\documentclass[prd,showkeys,preprint,11pt]{revtex4}
\usepackage{amsmath,amsthm,amsfonts,amssymb}
\usepackage[mathcal]{eucal}
\usepackage{mathrsfs}
\usepackage{graphicx}
\usepackage{dcolumn}
\usepackage{latexsym}
\usepackage{epsfig}
\usepackage{bm}
\usepackage[all]{xy}

\usepackage[english]{babel}
\usepackage{amsmath,amsthm,amsfonts,amssymb}
\usepackage{graphicx}
\usepackage{dsfont}
\usepackage{hyperref}

\DeclareMathOperator{\e}{e}
\DeclareMathOperator{\sech}{sech}
\DeclareMathOperator{\mx}{max}
\DeclareMathOperator{\mn}{min}

\newcommand{\ie}{\begin{equation}}
\newcommand{\fe}{\end{equation}}

\newcommand\fverb{\setbox\fverbbox=\hbox\bgroup\verb}
\newcommand\fverbdo{\egroup\medskip\noindent%
            \fbox{\unhbox\fverbbox}\ }
\newcommand\fverbit{\egroup\item[\fbox{\unhbox\fverbbox}]}
\newbox\fverbbox

\def\text#1{\mbox{#1}}

\usepackage{color}

\begin{document}
\title{Fermionic Kaluza-Klein modes in the string-cigar braneworld}
\author{D. M. Dantas$^{a}$}

\author{D. F. S. Veras$^{a}$}

\author{J. E. G. Silva$^{b}$}

\author{C. A. S. Almeida$^{a}$}

\address{$^{a}$Universidade Federal do Cear\'a (UFC), Departamento de F\'{\i}sica, Campus do Pici, Caixa Postal 6030, 60455-760, Fortaleza, Cear\'{a}, Brazil}

\address{$^{b}$Indiana University Center for Spacetime Symmetries, Bloomington, Indiana 47405, USA}



\begin{abstract}
We study the spin $1/2$ and spin $3/2$ fermion fields in a thick braneworld scenario in six dimensions called string-cigar model. This smooth string-like model has a source that satisfies the dominant energy condition and undergoes a Ricci flow. We propose a new coupling for the fermions with a background gauge field which allows a smooth and normalized massless mode in the brane with positive tension. By numerical methods the mass spectrum and the massive eigenfunctions are obtained. The Kaluza-Klein massive tower exhibits the usual increasing pattern and, in this scenario, the coupling term does not allow tachyonic Kaluza-Klein states. The brane core and the background gauge field alter the properties of the massive KK tower, enhancing the amplitude of the massive states near the origin and changing the properties of the analogue Schr\"{o}dinger potential. Furthermore, we find massive modes as resonant states in this scenario for both fermionic fields.



\end{abstract}


\maketitle


\section{Introduction}
\label{Sec_Introduction}

In the last years, the extra dimensions physics has acquired a conspicuous prominency, mainly due to the seminal models of large extra dimensions (ADD model) \cite{ADD} and the warped compactification models (RS model) \cite{rs,merab}. The RS model, for instance, brought to light again the possibility of noncompact extra dimension \cite{warpedmetrics}. These braneworld models provide solutions for important problems of the High Energy Physics, as the gauge hierarchy problem \cite{ADD,rs,merab}, the cosmological constant problem \cite{cosmologicalconstant} and the origin of dark matter \cite{darkmatter}.

Fields endowed with a bulk dynamics provide an explanation for the brane stability by assuming a stable source configuration, as a topological defect \cite{rsfields}. In 5D (or more generically in codimension $1$) models, a domain wall brane is generated by a spontaneous symmetry breaking mechanism which also provides a width and
inner dynamics to the brane \cite{rsfields}. Further, for bulk propagating fields reproduce the 4D physics, a massless Kaluza-Klein (KK) mode ought to have a compact support (localized) around the brane, what leads to a massless effective 4D action \cite{Kehagias:2000au}.
Nonetheless, unlike the gravitational and scalar fields, the vector gauge field has no localized massless mode \cite{Kehagias:2000au}, being required a coupling with the dilaton field \cite{Kehagias:2000au}. For the spin $1/2$ and $3/2$ fermions, the localization of the massless mode also require an additional coupling, usually a Yukawa coupling \cite{Csaki1,Csaki2,Cruz:2011kj, German:2012rv, Tofighi:2014vaa, Gauge-5D-Dilaton, CASA-Fermion-TwoField-ThickBrane, Chineses-Ressonance, Wilami1, Correa:2010zg, Castro:2010uj, Chumbes:2011zt}. The bulk dynamics and the additional coupling also provides the existence of KK massive resonant modes which brings important phenomenological consequences \cite{Csaki1,Csaki2,CASA-Fermion-TwoField-ThickBrane,Chineses-Ressonance}. Another approach to obtain a normalizable massless mode for the gauge and fermionic fields in brane models with positive tension is accomplished by changing of the geometry structure for a Weyl geometry \cite{weylgeometry}.

In six dimensions (or codimension 2), stationary braneworld scenarios with axial symmetry
are called string-like models \cite{stringlike,Gherghetta:2000qi}. The string-like branes inherited this name due their resemblance with topological defects in $(3+1)$, as the cosmic strings \cite{cosmicstring}.  Assuming a warped geometry with regularity conditions, the gravity is localized in the string-like models whose source is a global defect \cite{stringlike}, an infinitely thin brane \cite{Gherghetta:2000qi} or a local vortex \cite{Giovannini:2001hh}. The gauge vector field is also localized without any other coupling but the minimal gravitational coupling \cite{Oda:2000zc,Oda:2003jc,Oda-PRD}.

An important string-like scenario is the Gherghetta-Shaposhnikov (GS) model, whose bulk geometry is the warped product between a thin brane and the two dimensional disk \cite{Gherghetta:2000qi}. This vacuum solution of the Einstein equation with a negative cosmological constant traps gravity and provides a smaller
correction to the Newtonian potential compared to the RS model \cite{Gherghetta:2000qi}. Nonetheless, the GS model does not satisfy the regularities conditions at the origin and the dominant energy condition, as well \cite{tinyakov}.

Giovannini \textit{et al} found numerically a string-like solution wherein the source is a local Abelian vortex in the Einstein-Maxwell-Higgs model \cite{Giovannini:2001hh}. This smooth and thick string-like brane satisfies all the energy conditions and all the regularity conditions \cite{Giovannini:2001hh}. The components of the stress-energy tensor are concentrated around the origin for the winding number $n=0$. For higher winding numbers, the source is shifted from the origin \cite{Giovannini:2001hh}. Yet, only a numerical solution for this model is known.

String-like models with gravitational higher derivatives \cite{codimension2gravity} were proposed and their cosmological features analyzed \cite{codimension2cosmology}.  Braneworld models in 6D was also proposed assuming non-trivial transverse manifolds. Carlos and Moreno found a smooth string-like solution with a cigar-like shape \cite{cigaruniverse} whereas Kehagias proposed a conical tear drop-like transverse space to solve the cosmological constant problem \cite{teardrop}. Silva and Almeida used a section of the resolved conifold to build a thick string-like brane that
traps the gravitational \cite{conifold}, scalar and gauge fields \cite{Costa:2013eua}. Other transverse manifolds include general Einstein spaces \cite{einsteinmanifolds}, apple-shape \cite{apple}, football shape \cite{football}, and others \cite{smoothed}.

An analytical interior and exterior string-like model was proposed wherein the transverse manifold has a cigar-like shape, the so-called string-cigar model \cite{Silva:2012yj}.
The bulk geometry is a warped product between a $3$-brane and the so-called cigar soliton, a stationary solution of the Ricci which asymptotically converges to the 2D disk \cite{chow}. The Ricci flow appears in string theory as a RG flow of the worldsheet \cite{Friedan} and this geometrical flux has also applications topological massive gravity \cite{Lashkari:2010iy} and condensed matter \cite{Orth:2012ri}.

The cigar shape provides a metric similar to that found numerically by Giovannini \textit{et al} \cite{Giovannini:2001hh}, which makes the metric satisfy all the regularity conditions at the origin and behave similar to the GS model at large distances \cite{Silva:2012yj}.
Then, the string-cigar model can be regarded as a smooth extension of the GS model \cite{Silva:2012yj}. Since a Ricci soliton is an extension of the Einstein spaces, the string-cigar model generalizes the 6D model in Ref. \cite{einsteinmanifolds}. The Ricci flow in the transverse space makes the source undergo a flux that varies the brane-tensions and the bulk cosmological constant \cite{Silva:2012yj}. As a result, the source satisfies all the energy conditions and it has a bell-shaped core displaced from the origin, as in the Ref. \cite{Giovannini:2001hh}.


The spin $1$ and spin $2$ bulk fields have been already studied in the string-cigar brackground \cite{Graviton_Charuto, Costa:2015dva}. The massless modes are localized at the brane core and they recover the usual string-like behavior at large distances \cite{Gherghetta:2000qi,Oda-PRD}. The Kaluza-Klein spectra are attained, showing an increasing behavior for both fields, and the correspondent massive modes are enhanced near the brane core \cite{Graviton_Charuto, Costa:2015dva}. Besides, resonant massive states are present in the gravitational case \cite{Graviton_Charuto}.

A relevant question to address is how to include the matter fields (fermions) in the thin and thick string-like models. Besides the spin $1/2$ spinor which describes the ordinary matter, another important fermion is the spin $3/2$ gravitino field (superpartner of the graviton) that arises in supergravity context and is a dark matter candidate \cite{Panotopoulos:2007fg}.  Oda studied the localization of various spin fields in a thin string-like scenario \cite{Oda-PRD, Oda:2000zc}.
As the vector gauge field in 5D, the free spin $1/2$ and $3/2$ fermions can only be localized (without additional interactions) on a string-like defect with the exponentially  increasing warp factor (negative tension) \cite{Oda:2000zc}. Xiao Liu \textit{et al} proposed to couple the fermions with a $U(1)$ background gauge field by which the zero mode is confined in the thin string-like brane with positive tension \cite{Liu:2007gk}. Parameswaran \textit{et al} analyzed the massless and massive spectrum in a 6D supergravity model which enables a finite mass gap even for an infinite extra dimension \cite{Parameswaran:2006db}. Dantas \textit{et al} \cite{Dantas:2013iha} obtained a normalized massless mode for a fermion in a string-like brane with a transverse resolved conifold \cite{conifold}. The depth of the potential well and the high of the potential barrier evolves with the resolution parameter \cite{Dantas:2013iha}. However, although normalizable, the massless mode is not well-defined at the origin.


In this work, we propose a new coupling for the fermions with a background gauge field which localizes the zero mode in the string-cigar model. The thin-string limit is considered, as well. Imposing suitable boundary conditions to guarantee the self-adjointness of the spinor operators, a normalized and everywhere well-defined massless mode is obtained for both the thin string and string-cigar models. As for the gravitational \cite{Graviton_Charuto} and the $U(1)$ \cite{Costa:2015dva} vector fields, we find that the spin $1/2$ field massless mode is shifted from the origin and sets around the displaced brane core. The mass spectrum is the same for both right-handed and left-handed chiralities and it is free of tachyons. Further, the KK spectrum has an increasing pattern which exhibits the usual linear behaviour of the Kaluza-Klein theories. For the Rarita-Schwinger field (spin $3/2$), the zero mode and massive spectrum have minor changes when compared to those of the spin $1/2$. However, the amplitudes of the massive eigenfunctions for the spin $3/2$ are higher than those for the spin $1/2$. In comparison with the thin string-like model, the core of the string-cigar brane enhances the massive modes near the origin for both spin $1/2$ and $3/2$ fields. Besides, in a Schr\"{o}dinger approach, the spin $1/2$ and $3/2$ fields possess identical behaviour. Furthermore, the coupling also enables the presence of resonant modes (massive modes solutions having very large amplitude near the brane).

This paper is organized as follows: in section \ref{Sec_StringCigarBraneword}, we review the features of the thin string and the string-cigar models.
We comment on the main properties of the gravity and gauge fields in these models, which were developed in References  \cite{Graviton_Charuto} and \cite{Costa:2015dva}, respectively. We also studied the dynamics of the scalar field in these scenarios and show that is it identical to the gravitational case \cite{Graviton_Charuto}.
In sections \ref{Sec_Fermion-1/2} and \ref{Sec_Fermion-3/2}, we study the localization of spin $1/2$ and $3/2$ fields. Furthermore, the massive spectrum is obtained in both cases. Concerning the resonant modes, we conclude that the Schr\"{o}dinger-like potentials are the same for both spin $1/2$ and spin $3/2$ cases. Thus the behaviour of the resonant modes are only studied in subsection \ref{Sec_Ress}.  In section \ref{conclusions}, conclusions and perspectives are outlined.


\section{The string-cigar braneworld}
\label{Sec_StringCigarBraneword}

Consider a six dimensional spacetime $\mathcal{M}_{6}$ where the 3-brane $\mathcal{M}_{4}$ can be embedded.
Assuming an axially symmetric and static bulk $\mathcal{M}_{6}$, the braneworld scenario is called a string-like model \cite{stringlike,gregory,Gherghetta:2000qi,Giovannini:2001hh}.
A general metric for the string-like model takes the form \cite{stringlike,gregory,Gherghetta:2000qi,Giovannini:2001hh}
\begin{eqnarray}\label{metric}
ds^2_6=F(r)\eta_{\mu\nu}dx^{\mu}dx^{\nu}+dr^2+H(r)d\theta^2 ,
\end{eqnarray}
where $0\leq r\leq r_{\mx}$ and $\theta\in[0,2\pi)$ are the radial and the angular coordinates, respectively. The radial component can extend to infinity, i.e., $r_{\mx}\rightarrow\infty$.
In order to guarantee that the scalar and the extrinsic curvatures are finite at the origin, it is usual to impose the
regularity conditions \cite{stringlike,Gherghetta:2000qi,Giovannini:2001hh}
\begin{equation}
F(0) = \left(\sqrt{H(r)}\right)^{\prime}_{r=0} = 1, \quad \text{and} \quad  F^{\prime}(0) = H(0) = 0,
\label{regularity}
\end{equation}
where the primes denote derivatives with respect to $r$.

Performing the change of coordinate \cite{ponton}
\begin{equation}
z(r)=\int_{0}^{r}{F^{-\frac{1}{2}}(r')dr^{\prime}},
\label{Eq_z(r)}
\end{equation}
the metric \eqref{metric} can be cast in the conformal form \cite{ponton}
\begin{equation}
ds^2_6=F(z)\left(\eta_{\mu\nu}dx^{\mu}dx^{\nu} + dz^2 + \beta(z)d\theta^2\right),
\end{equation}
where $\beta(z):=\frac{H(z)}{F(z)}$.

Let us assume that the bulk dynamic is governed by the Einstein-Hilbert action with bulk cosmological constant $\Lambda$ \cite{stringlike,gregory,Gherghetta:2000qi,Giovannini:2001hh}:
\begin{equation}
\label{einsteinhilbertaction}
  S_{g} =\int_{\mathcal{M}_{6}}{\left(\frac{1}{2\kappa_{6}}R-\Lambda +\mathcal{L}_{m}\right)\sqrt{-g}d^{6}x},
\end{equation}
where $\kappa_{6}=8\pi/M_{6}^{4}$, $M_{6}^{4}$ is the six-dimensional bulk Planck mass and $\mathcal{L}_{m}$ is the matter Lagrangian for the source of the geometry. From the matter Lagrangian $\mathcal{L}_{m}$ we define the stress-energy tensor
\begin{equation}
\mathbf{T}=T_{MN}dx^{M}\otimes dx^{N},
\end{equation}
where the stress-energy tensor components are defined by \cite{stringlike,gregory,Gherghetta:2000qi,Giovannini:2001hh}.
\begin{equation}
T_{MN}=\frac{2}{\sqrt{-g}}\frac{\partial (\sqrt{-g} \, \mathcal{L}_{m})}{\partial g^{MN}}.
\end{equation}
The Einstein-Hilbert action provides the Einstein equation for the bulk \cite{stringlike,gregory,Gherghetta:2000qi,Giovannini:2001hh}:
\begin{equation}
R_{MN} - \frac{1}{2}R g_{MN} = -\kappa_6 (\Lambda g_{MN} + T_{MN}).
\label{Eintein-Equations}
\end{equation}
An axisymmetric and static anstaz for the stress-energy tensor has the form \cite{stringlike,gregory,Gherghetta:2000qi,Giovannini:2001hh}.
\begin{equation}
\label{stressenergyansatz}
\mathbf{T}=t_{0}(r)\left(\mathbf{e}_{0}\otimes \mathbf{e}_{0} + \sum_{i=1}^{3}\mathbf{e}_{i}\otimes\mathbf{e}_{i}\right) + t_{r}(r)\mathbf{e}_{r}\otimes \mathbf{e}_{r} + t_{\phi}(r)\mathbf{e}_{\phi}\otimes \mathbf{e}_{\phi}.
\end{equation}
Using the metric ansatz \eqref{metric} and the stress-energy ansatz \eqref{stressenergyansatz}, the bulk Einstein equation 
\eqref{Eintein-Equations} yields to the system of equations
\begin{eqnarray}
\label{einsteinequationsystem}
 \frac{3}{2}\left(\frac{F'}{F}\right)' + \frac{3}{2}\left(\frac{F'}{F}\right)^{2} + \frac{3}{4}\frac{F'}{F}\frac{H'}{H} + \frac{1}{4}\left(\frac{H'}{H}
\right)^ { 2 } +\frac {1} {2} \left(\frac { H'}{H}\right)' &  =   &   -\kappa_{6}(\Lambda+t_{0}(r)),\\
\frac{3}{2}\left(\frac{F'}{F}\right)^{2}+\frac{F'}{F}\frac{H'}{H} &   =   &   -\kappa_{6}(\Lambda+t_{\rho}(r)), \\
2\left(\frac{F'}{F}\right)'  +  \frac{5}{2}\left(\frac{F'}{F}\right)^{2}    &   =   &    -\kappa_{6}(\Lambda+t_{\theta}(r)),
\end{eqnarray}

\subsection{Thin string-like model}
In the Gherghetta-Shaposhnikov (GS) model, a vacuum solution of the system of Einstein equations \eqref{einsteinequationsystem} was found \cite{Gherghetta:2000qi}. Assuming that $\frac{F'}{F}=\frac{H'}{H}=0$, the metric functions are given by \cite{Gherghetta:2000qi}
\begin{eqnarray}
\label{gsmetric}
F(r)=\e^{-cr}	&	,	&	H(r)=R_{0}^{2}F(r),
\end{eqnarray}
where $R_{0}$ is an arbitrary length scale and the constant $c$ is related to the bulk cosmological constant by \cite{Gherghetta:2000qi}
\begin{equation}
c^{2}=-\frac{2}{5}\kappa_{6}\Lambda.
\end{equation}
Hence, the cosmological constant must be negative and the bulk is an $AdS_{6}$ spacetime \cite{Gherghetta:2000qi}. Since the functions in Eq. \eqref{gsmetric} are vacuum solutions, the GS model represents an infinitely thin string-like braneworld which is an extension of the Randall-Sundrum (RS) metric to six dimensions.

For the thin string-like metric, i.e., $F(r)=\e^{-cr}$ and $H(r)=R_{0}^2 F(r)$, the conformal coordinate can be found to be \cite{ponton}
\begin{equation}
\label{thinstringconformalcoordinate}
z(r)=\frac{2}{c}\left(\e^{\frac{cr}{2}}-1\right).
\end{equation}
Note that $z(r=0)=0$ and $z'(r)>0$ in Eq. \eqref{thinstringconformalcoordinate}. Then, the conformal coordinate $z(r)$
is still a gaussian radial coordinate which measures the distance from a point in the transverse manifold to the origin.
Using the conformal coordinate $z$ in Eq. \eqref{thinstringconformalcoordinate}, the metric factors have the form
\begin{equation}
F(z)=\frac{4}{c^{2}}\frac{1}{\left(z+\frac{2}{c}\right)^{2}} \quad \text{,} \quad H(z)=R_0^2F(z).
\end{equation}

Furthermore, the hierarchy problem between the four-dimensional Planck mass ($M^4$) and the
bulk Planck mass ($M^6$) is solved in this scenario   \cite{Gherghetta:2000qi, Silva:2012yj}, and these masses are related by the following equation
\begin{eqnarray}
M^2_P	&	=	&	2\pi M_6^4\int_{r'=0}^{r_{\mx}}{\sqrt{-g(r)} \, F^{-1}(r')dr'}\nonumber\\
		&	=	&	\frac{2\pi R_{0}}{3c}\left(1-\e^{\frac{3 c r_{\mx}}{2}}\right).
\label{hierarchy}
\end{eqnarray}
The relation between the bulk and brane Planck energies can be rewritten using only the ratio between the bulk cosmological constant and the string tension \cite{Gherghetta:2000qi}.

For fluctuations of the metric \eqref{metric} in the form \cite{Gherghetta:2000qi,Giovannini:2001hh,Silva:2012yj}
\begin{equation}
ds_6^2 = F(r)(\eta_{\mu\nu} + h_{\mu\nu})dx^{\mu}dx^{\nu} + dr^2 + H(r)d\theta^2,
\label{Gravitation-Perturbation}
\end{equation}
satisfying the traceless gauge $\nabla^{\mu}h_{\mu\nu} = 0$, the linearization of the Einstein equations \eqref{Eintein-Equations}  yields the equation for the gravitational perturbation \cite{Gherghetta:2000qi, Silva:2012yj}
\begin{equation}
\Box_6 h_{\mu\nu} = \partial_A ( \sqrt{-g_6} \: \eta^{AB} \partial_B h_{\mu\nu} ) = 0.
\label{Graviton-Equation}
\end{equation}
Performing the Kaluza-Klein decomposition $h_{\mu\nu}({\bf x},r,\theta) = \tilde{h}_{\mu\nu}({\bf x})\sum_{n, l=0}^{\infty}\phi_{n,l}(r)\e^{il\theta}$ \cite{Gherghetta:2000qi, Silva:2012yj} and imposing the free wave dependence on the $3-$brane $\Box_4\tilde{h}_{\mu\nu}({\bf x}) = m^2	\tilde{h}_{\mu\nu}({\bf x})$ \cite{Gherghetta:2000qi,Giovannini:2001hh,Silva:2012yj}, the radial component of the graviton equation of motion takes the general form \cite{Gherghetta:2000qi,Giovannini:2001hh,Silva:2012yj}
\begin{equation}
\left[ \partial_r^2 + 2\mathcal{P} \: \partial_r + \left( \frac{m_{n,l}^2}{F} - \frac{l^2}{H}\right) \right]\phi_{n, l}(r) = 0,
\label{Sturm-Liouville_Graviton}
\end{equation}
where
\begin{equation}
\mathcal{P}(r) = \frac{F^{\prime}}{F} + \frac{1}{4}\frac{H^{\prime}}{H} = \frac{5}{4}\frac{F^{\prime}}{F} + \frac{1}{4} \frac{\beta^{\prime}}{\beta}.
\label{P(r)}
\end{equation}
The graviton radial equation for the thin string-like model has the explicit form \cite{Gherghetta:2000qi}
\begin{equation}
\phi_m^{\prime\prime} - \dfrac{5}{2}c\phi_m^{\prime} +\left(m_0^2 - l^2/R_0^2\right)\e^{cr}\phi_m = 0.
\label{Sturm-Liouville-GS-graviton}
\end{equation}
From the Eq. \eqref{regularity}, we impose the boundary conditions \cite{Gherghetta:2000qi,Giovannini:2001hh,Silva:2012yj}
\begin{eqnarray}\label{cc2}
\phi^{\prime}(\infty)= \phi^{\prime}(\infty)=0.
\end{eqnarray}
The radial graviton equation \eqref{Sturm-Liouville-GS-graviton} together with the boundary conditions \eqref{cc2} forms a
Sturm-Liouville problem \cite{Gherghetta:2000qi,Giovannini:2001hh,Silva:2012yj}. For $m = 0$ and s-wave solution ($l=0$), the localized massless mode solution is obtained as \cite{Gherghetta:2000qi}
\begin{equation}
\tilde{\phi}_{m = 0}(r) = \sqrt{\frac{3c}{2R_0}}\e^{-\frac{3}{4}cr},
\label{Massless-Mode_GS}
\end{equation}
where $\tilde{\phi}_m =  \e^{-\frac{3}{4}cr}\phi_{m = 0}$ \cite{Gherghetta:2000qi}.

For $m\neq 0$, the massive modes has the form \cite{Gherghetta:2000qi}
\begin{equation}
\phi_m(\rho) = \e^{\frac{5}{4} c\rho}\left[ B_1 J_{5/2}\left(\frac{2m}{c}\e^{\frac{1}{2}c\rho}\right) + B_2 Y_{5/2}\left(\frac{2m}{c}\e^{\frac{1}{2}c\rho}\right)\right],
\label{GS-MassiveModes}
\end{equation}
where $B_1$ and $B_2$ are arbitrary constants and $m = m_0^2 - l^2/R_0^2$. The exponential dependence reveals that the massive modes are not localized on the brane \cite{Gherghetta:2000qi}. Applying the boundary conditions \eqref{cc2} on the massive modes \eqref{GS-MassiveModes}, the graviton mass spectrum in the GS model was found as \cite{Gherghetta:2000qi}:
\begin{equation}
m_n \simeq c \left( n - \frac{1}{2} \right)\frac{\pi}{2}\e^{-\frac{c}{2}r_{\mx}},
\label{Spectrum_GS}
\end{equation}
where $r_{\mx}$ is a finite radial distance cutoff.
The gravitational massless mode is localized in the brane and the contribution from the nonzero modes provides a small correction to the Newton’s law on the $3-$brane  \cite{Gherghetta:2000qi}.

The vector gauge field was also studied in the string-like models \cite{Oda:2000zc,Oda:2003jc,Oda-PRD,Parameswaran:2006db,Giovannini_Gauge-6D-CQG,Giovannini_Gauge-6D-PRD,Costa:2013eua,Costa:2015dva}. Starting with action 
\begin{eqnarray}\label{action-gauge-6D}
 S_{spin-1}=\int{\sqrt{-g} \, g^{MN}g^{RS}\mathcal{F}_{MN}\mathcal{F}_{RS}}d^6x,
\end{eqnarray}
where $\mathcal{F}_{MN}=\nabla_{M}\mathcal{A}_{N}-\nabla_{N}\mathcal{A}_M$, the equation of motion is \cite{Oda:2000zc,Oda:2003jc,Oda-PRD}
\begin{eqnarray}\label{eqm-gauge-6D}
 \frac{1}{\sqrt{-g}}\left(\partial_{S}\sqrt{-g} \, g^{SM}g^{RN}\mathcal{F}_{MN}\right)=0.
\end{eqnarray}
Imposing the gauge conditions \cite{Oda:2000zc,Oda:2003jc,Oda-PRD} $\partial_{\mu}\mathcal{A}^{\mu}=\mathcal{A}_{\theta}=0$ and $\mathcal{A}_{r}=\mathcal{A}_r(r, \theta)$, the Maxwell equations read \cite{Oda-PRD, Oda:2000zc, Costa:2013eua, Costa:2015dva}
\begin{eqnarray}\label{eqm-gauge2}
\left(\eta^{\mu \nu}\partial_{\mu}\partial_{\nu}+\frac{F}{H}\partial_{\theta}^2\right)\mathcal{A}_{r}=0,\\
\label{eqm-gauge3}
\partial_{r}\left(\frac{F^2}{\sqrt{H}}\partial_{\theta}\mathcal{A}_{r}\right)=0,
\\
\label{eqm-gauge4}
\left(\eta^{\mu \nu}\partial_{\mu}\partial_{\nu}+\frac{F}{H}\partial_{\theta}^2+\frac{1}{\sqrt{H}}\partial_{r}\left(F\sqrt{H}\partial_{r}\right)\right)\mathcal{A}_{\lambda}=0.
\end{eqnarray}
Using the Kaluza-Klein decompositions $\mathcal{A}_{\mu}(x^{M},r,\theta)=\sum\limits_{n, l=0}^{\infty}\mathcal{A}_{\mu}^{(n,l)}(x^\mu)\rho_n(r)\e^{il\theta}$ and  $\mathcal{A}_{r}(x^{M},r,\theta)=\sum\limits_{l=0}^{\infty}\mathcal{A}_{\mu}^{(l)}(x^\mu)\varrho(r)\e^{il\theta}$,
the radial dependence of the gauge field on the brane for $l=0$ is governed by the Sturm-Liouville equation \cite{Costa:2013eua, Costa:2015dva}
\begin{eqnarray}\label{sl-spin1}
\left[\partial_r^2+ \left(2\mathcal{P}-\frac{F^{\prime}}{F}\right)\partial_r + \frac{m_n^2}{F}\right]\rho_n(r)=0,
\end{eqnarray}
which can be rewritten as
 \begin{equation}
 \label{chiequation}
 \rho_n''(r)+\left(\frac{3}{2}\frac{F'}{F}+\frac{1}{2}\frac{{\beta}'}{\beta}\right)\rho_n'(r)+\frac{m_n^2}{F}\rho_n(r)=0,
 \end{equation}
where $\beta(r,c) = H(r,c)/F(r,c)$. For the thin-string model, the radial equation \eqref{chiequation} has the explicit form \cite{Oda-PRD, Oda:2000zc}
\begin{equation}
\phi_m^{\prime\prime} - \dfrac{3}{2}c\phi_m^{\prime} +\left(m_0^2 - l^2/R_0^2\right)\e^{c\rho}\rho_m = 0.
\label{Sturm-Liouville-GS-gauge}
\end{equation}
Unlike the vector gauge field in the RS model, in the thin string-like model the gauge field has a massless mode of form \cite{Oda-PRD, Oda:2000zc}
\begin{equation}
\label{thinmasslessmode}
\rho_{0}(r)=\sqrt{\frac{5c}{2R_{0}}}\e^{-\frac{c}{2}r},
\end{equation}
which is normalizable \cite{Oda-PRD, Oda:2000zc}. The massive modes have the form \cite{Oda-PRD, Oda:2000zc}
\begin{equation}
\rho_m(r) = \e^{\frac{3}{4} cr}\left[ B_1 J_{3/2}\left(\frac{2m}{c}\e^{\frac{1}{2}cr}\right) + B_2 Y_{3/2}\left(\frac{2m}{c}\e^{\frac{1}{2}cr}\right)\right],
\label{GS-MassiveModes-vector}
\end{equation}
and then, are non-localizable \cite{Oda-PRD, Oda:2000zc}. Imposing the same boundary conditions \eqref{cc2}, a linearly increasing KK gauge spectrum was found \cite{Oda-PRD, Oda:2000zc}
\begin{equation}
m_n = \frac{c}{2}n\pi\e^{-\frac{c}{2}r_{\max}}.
\label{Spectrum-Gauge}
\end{equation}

Finally, for a minimally coupled scalar field, the action \cite{Oda-PRD, Oda:2000zc,conifold}
\begin{eqnarray}\label{action-spin0}
 S_{spin-0}=-\frac{1}{2}\int{\sqrt{-g} \, g^{MN}\partial_{M}\Phi\partial_{N}\Phi}d^6x,
\end{eqnarray}
provides the equation of motion 
\begin{eqnarray}\label{eqm-spin0}
\frac{1}{\sqrt{-g}}\partial_M\left[\sqrt{-g} \, g^{MN}\partial_{N}\Phi\right]=0.
\end{eqnarray}
Using the general string-like metric \eqref{metric} in Eq. \eqref{eqm-spin0}, we have
\begin{eqnarray}\label{eqm-spin0b}
\left[\frac{\eta^{\mu \nu}}{F}\partial_{\mu}\partial_{\nu}+\frac{\partial_r\left(F^2H^{\frac{1}{2}}\partial_r\right)}{F^{2}H^{\frac{1}{2}}}+\frac{\partial^2_{\theta}}{H}\right]\Phi=0.
\end{eqnarray}
Using the KK decomposition $\Phi(x^{\mu}, r, \theta)=\frac{1}{\sqrt{\pi}}\varphi(x^{\mu})\sum\limits_{n, l = 0}^{\infty}\chi_n(r)\e^{il\theta}$ and $\eta^{\mu \nu}\partial_{\mu}\partial_{\nu}\varphi(x^{\mu})=m^2\varphi(x^{\mu})$, the Eq. \eqref{eqm-spin0} yields the radial sturm-liouville equation
\begin{eqnarray}\label{sl-spin0}
\left[\partial_r^2+ 2\mathcal{P}(r)\partial_r + \left(\frac{m^2_{n,l}}{F}-\frac{l^2}{H}\right)\right]\chi_n(r)=0.
\end{eqnarray}
Note that this equation is the same for the gravitational case \eqref{Sturm-Liouville_Graviton}, thus  for s-wave solution, both zero mode and as the KK spectrum are given by the results of the graviton \cite{Oda:2000zc,conifold,Silva:2012yj,Graviton_Charuto}. This  behavior between Spin $0$ and spin $2$ modes is also verified in five dimensional scenarios \cite{Gauge-5D}.

Despite the good results of the thin string-like model described above, the regularity conditions \eqref{regularity} at the origin and the dominant energy conditions are not satisfied for this model \cite{Giovannini:2001hh,tinyakov}. This issue arises due to the metric be only an exterior solution of the Einstein equations. Although it is possible to consider an interior and exterior solution separately using the junctions conditions to match the solutions, the thin string limit (where the width of the core vanishes and only the exterior solution remains) is known to present the same problems \cite{tinyakov}. As a matter of fact, Giovannini \textit{et al} have found numerically an interior and exterior string-like solution of the Einstein equations
satisfying all the regularity and energy conditons whose source is an Abelian vortex in the Einstein-Maxwell-Higgs model. Nonetheless, the analytical expression of the solution is yet unknown.

\subsection{String-cigar model}
An analytical interior and exterior smooth extension of the GS model, called the string-cigar model, was proposed in Ref. \cite{Silva:2012yj}. The thin string-like GS model is built from the warped product between the $3-$brane and two dimensional disc $\mathbb{D}^{2}$ of radius $R_{0}$, whose metric has the form
\begin{equation}
\label{discmetric}
ds^{2}_{\mathbb{D}^2}=dr^2 + R_{0}^2 d\theta^2.
\end{equation}
The constancy of the radius prevent the GS model to satisfy the $H(0)=0$ and $(\sqrt{H})'=1$ conditions.

In the Ref. \cite{Silva:2012yj}, a regular transverse manifold $\mathcal{C}_{2}$ with metric
\begin{equation}
\label{cigarmetric}
ds^2_{\mathcal{C}_{2}}=dr^2 + \frac{1}{c^2}\tanh^2(cr)d\theta^2
\end{equation}
was proposed. Note that asymptotically, the metric \eqref{cigarmetric} converges to the $\mathbb{D}^2$ metric \eqref{discmetric},
for $R_{0}=1/c$. At the origin, the effective radius $\tanh(cr)/c$ vanishes. Hence, the transverse manifold $\mathcal{C}_2$ has a cigar-like behaviour.

\begin{figure}[htb]
 \centering
    \includegraphics[width=0.50\textwidth]{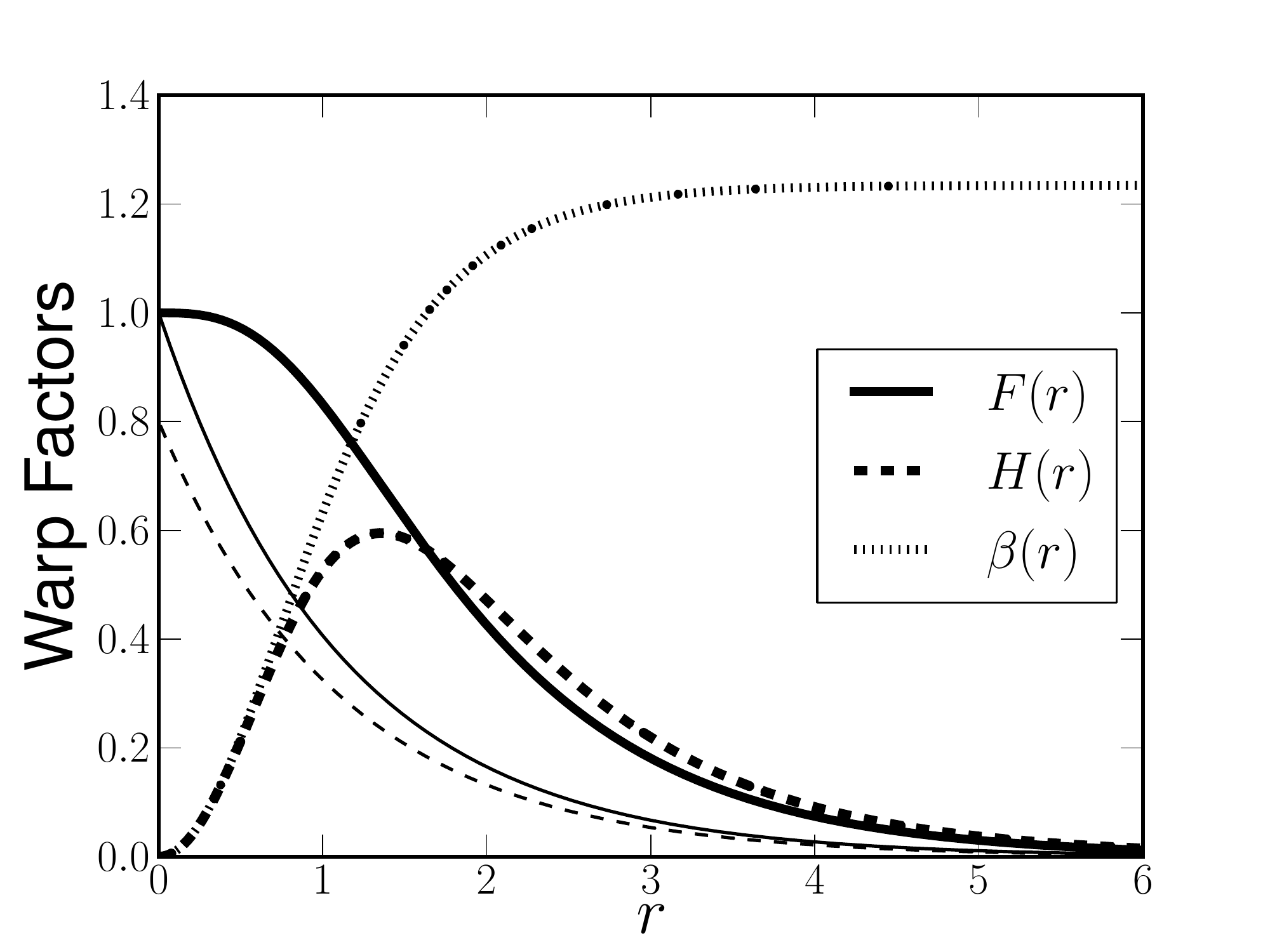}
 \caption{Warp factors for $c = 0.9$. The thick lines correspond to the  string-cigar geometry, while the thin lines to the GS model with $R_0 = 8.0$.}
 \label{Fig_WarpFactors}
\end{figure}

Indeed, the transverse space $\mathcal{C}_2$ is a stationary solution of the geometric Ricci flow according to \cite{chow, Friedan,Silva:2012yj}
\begin{equation}
\frac{\partial g_{ab}}{\partial c} = -2R_{ab}(c),
\label{Ricci_Flow}
\end{equation}
where $c$ can be regarded as a metric parameter and $R_{ab}$ is the Ricci tensor. The metric solution \eqref{cigarmetric} is
a Ricci flow solution called cigar soliton \cite{chow,Silva:2012yj}. The Ricci flow \eqref{Ricci_Flow} defines a family of smooth geometries depending on the parameter $c$ \cite{Silva:2012yj}.
Using the cigar soliton $\mathcal{C}_2$ as the transverse manifold, a string-like model was proposed as \cite{Silva:2012yj}
\begin{equation}
F(r)=\e^{-[cr -\tanh{(cr)}]},
\label{Eq_WarpFactors}
\end{equation}
and
\begin{equation}
 H(r)=\beta(r)F(r),
\end{equation}
where $\beta(r) = \frac{1}{c^2}\tanh^2{(cr)}$. Since this string-like mode has a cigar transverse manifold, it was named string-cigar model. We plot in the Figure \ref{Fig_WarpFactors} the metric functions for the thin string-like and for the string-cigar models. It can be seen that, for large $r$, both models possess similar behaviours. Therefore, the string-cigar model recovers the thin string-like limit asymptotically.
At the origin, the string-cigar model satisfies all the regularity conditions \cite{Silva:2012yj}.

\begin{figure}[htb]
 \centering
    \includegraphics[width=0.50\textwidth]{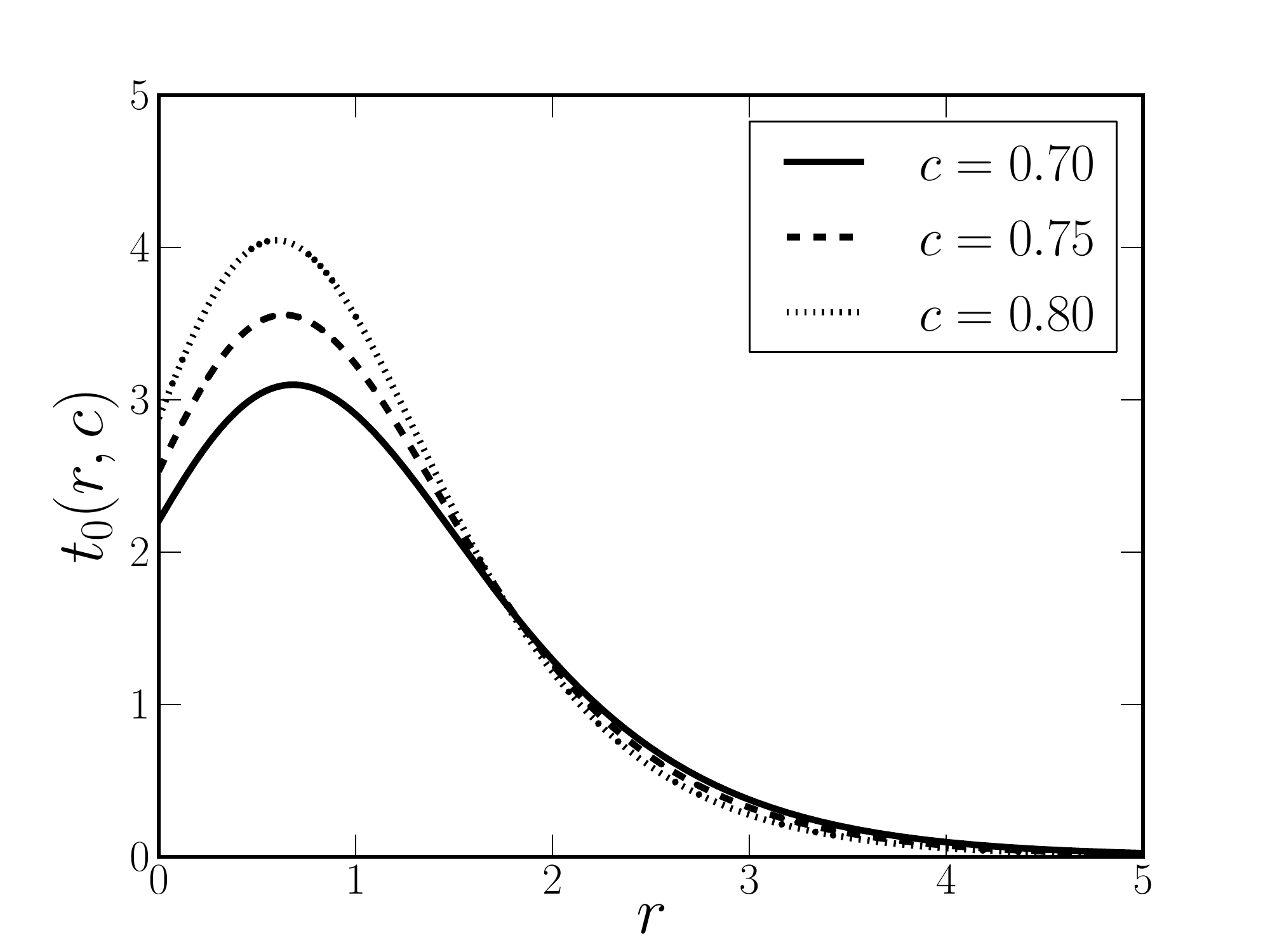}
 \caption{Energy density of the String-Cigar braneworld for different values of $c$. Its maximum indicates the position of the core of the brane.}
 \label{Fig_Energy-Density}
\end{figure}

From the Einstein equations \eqref{einsteinequationsystem}, the components of the stress-energy tensor corresponding to the string-cigar braneworld are \cite{Silva:2012yj}
\begin{eqnarray}
\label{energymomentumtensor}
t_{0}(r,c) & = & \frac{c^{2}}{\kappa_6}\Big( 7\text{sech}^{2}{cr}+\frac{13}{2}\text{sech}^{2}{cr}\tanh{cr}-\frac{5}{2}\text{sech}^{4}{cr}\Big)\\
 t_{r}(r,c) & = & \frac{c^{2}}{\kappa_{6}}\Big( 5\text{sech}^{2}{cr}+ 2\text{sech}^{2}{cr}\tanh{cr} - \frac{5}{2}\text{sech}^{4}{cr}\Big)\\
 t_{\theta}(r,c) & = & \frac{c^{2}}{\kappa_{6}}\Big(5\text{sech}^{2}{cr}+4\text{sech}^{2}{cr}\tanh{cr}-\frac{5}{2}\text{sech}^{4}{cr}\Big).
\end{eqnarray}
The components are non-negative $t_0,t_r,t_\theta \geq 0$ and satisfy the dominant energy condition, i.e., $t_0\geq |t_r|,|t_\theta|$.
We plot in the Figure \ref{Fig_Energy-Density} the energy density $t_0$ for different values of the parameter $c$. The position of the maximum of the energy-density evinces that the brane core is shifted from the origin, as for the Abelian vortex model for higher winding number \cite{Giovannini:2001hh}.

Since the stress-energy tensor components have a compact support around the origin, the string-cigar geometry can be
regarded as an interior and exterior thick string-like solution of the Einstein equations \eqref{einsteinequationsystem}. Once the
source evolves with the Ricci flow parameter $c$, the string-cigar model reflects changes that both the brane source and the bulk cosmological constant undergo \cite{Silva:2012yj}. For instance, due to the $\tanh(cr)$ and $\text{sech}(cr)$ terms, the stress-energy components vanish when $c\rightarrow\infty$ and we recover the thin GS model.

In the string-cigar model, the relation between the bulk and the brane energy scales, given by Eq. \eqref{hierarchy}, reads \cite{Silva:2012yj}
\begin{equation}\label{i-hierarchy}
M_4^2=\frac{2\pi M_6^4}{c}\int_{0}^{\infty}\frac{\e^{-\frac{3}{2}\left(cr -\tanh(cr)\right)}}{\tanh(cr)}dr.
\end{equation}
Then, in order to guarantee that $M_{4} \gg M_{6}$, the parameter $c$ must be small \cite{Silva:2012yj}.

The localization of gravity and $U(1)$ vector gauge field in the string-cigar braneworld was performed in References \cite{Silva:2012yj} and \cite{Costa:2015dva}, respectively.

The radial graviton equation has the form \cite{Silva:2012yj}
\begin{equation}
\label{massivemodeequation2}
\phi_m''+c\left[-\frac{5}{2}\tanh^{2}{(c\rho)} +  \hspace{0.1cm} \frac{\text{sech}^{2}(c\rho)}{\tanh{(c\rho})}\right]\phi_m'+\e^{(c\rho-\tanh{(c\rho)})}\left(m_n^{2}-\frac{l^{2}c^{2}}{
\tanh^ { 2 } {(c\rho)}} \right)\phi_m = 0.
\end{equation}
Note that asymptotically, $\tanh(cr)=1$ and $\text{sech}(cr)=0$, then the graviton equation in the string-cigar model
has the same form of the thin string-like GS model \cite{Silva:2012yj}. Nevertheless, near the origin, the effects of the
brane core change the behaviour of the gravitons. In fact, its massless mode is given by \cite{Silva:2012yj}
\begin{equation}
\tilde{\phi}_{m = 0}(r) = N_1 F^{\frac{3}{4}}(r)\beta^{\frac{1}{8}}(r),
\label{Massless_Grav-Charuto}
\end{equation}
where $N_1$ is a normalization constant. The massless graviton mode \eqref{Massless_Grav-Charuto} satisfies the boundary conditions \eqref{cc2} and asymptotically behaves as the GS massless mode. Further, the zero mode has its peak shifted from the origin, as the energy density \cite{Silva:2012yj}.


The massive modes as solutions of the Eq. \eqref{Sturm-Liouville_Graviton} was studied in details in Ref. \cite{Graviton_Charuto}, where the complete Kaluza-Klein spectrum and the corresponding eigenfunctions were attained. Asymptotically, the massive KK states has a similar behavior of the thin GS model whilst near the origin, the brane core interaction enhance their amplitude compared with in the GS model. The resonant states were also found as solutions of the analogue Schr\"{o}dinger equation \cite{Graviton_Charuto}. In the string-cigar model, the graviton massive spectrum, obtained numerically, has a linear behaviour, as for the GS model \cite{Graviton_Charuto}.

Since the radial equation for the scalar field is the same of the graviton, the analysis of the massive modes for the scalar field in the string-cigar braneworld is identical to the gravitational field presented in Ref. \cite{Graviton_Charuto}.

For the vector gauge field, the radial equation reads \cite{Costa:2015dva}
\begin{equation}
\label{CompleteEquationChi}
\rho_n'' + c\left[-\frac{3}{2}\tanh^{2}cr+ \frac{\sech^{2}cr}{\tanh cr}\right]\rho_n' + \e^{(cr-\tanh{cr})}m_n^2\rho_n=0.
\end{equation}
Asymptotically, the gauge KK equation \eqref{CompleteEquationChi} has the same form of that for the thin string-like model.
The localized massless solution of the Eq. \eqref{CompleteEquationChi} in the string-cigar scenario were found in Ref. \cite{Costa:2015dva} as
\begin{equation}
\tilde{\rho}_{m=0}(r) = N_2 F^{\frac{1}{2}}(r)\beta^{\frac{1}{4}}(r),
\label{Massless_Gauge-Charuto}
\end{equation}
where $N_2$ is a normalization constant. It is interesting to compare the massless mode of the bosonic and fermionic fields. We insert a comparative plot of massless modes of gravity, Eq.\eqref{Massless_Grav-Charuto}, vector field , \eqref{Massless_Gauge-Charuto} and fermionic fields in the Sec. \ref{Sec_Fermion-3/2}.

The massive solutions of Eq. \eqref{sl-spin1} for the string-cigar was studied in Ref. \cite{Costa:2015dva}. The well-known linear increasing behavior for $m \ll c$ was obtained.
As for the gravitational case \cite{Graviton_Charuto}, the massive eigenfunctions behave as the thin string-like modes asymptotically and they are influenced by the core of the brane near the origin. However, resonant states were not found \cite{Costa:2015dva}.

The similarity between Kaluza-Klein modes of the spin $0$ and spin $2$ fields is expected. This behaviour is present in the thin string-like model as inferred by Oda in Ref. \cite{Oda:2000zc}. Besides, the same feature occurs for the of spin $1/2$ and spin $3/2$ cases, which will be discussed in the sections \ref{Sec_Fermion-1/2} and \ref{Sec_Fermion-3/2}.


\section{Spin $1/2$ Fermions} 
\label{Sec_Fermion-1/2}

Consider the action on curved space background for bulk massless spin $\frac{1}{2}$ fermions  \cite{Oda:2000zc,Liu:2007gk, Liu:2007eb, Parameswaran:2006db}:
\begin{eqnarray}
\label{action-fermions-6D}
 S_{6_{1/2}}=\int{\sqrt{-g}\bar{\Psi} i\Gamma^{M} D_M \Psi}d^6x,
\end{eqnarray}
where $\Gamma^{M}=\xi^{M}_{\bar{M}}\Gamma^{\bar{M}}$ are the curved Dirac matrices defined from the flat Dirac matrices $\Gamma^{\bar{M}}$ through the vielbeins  $g_{MN}=\xi^{\bar{M}}_{M}\xi^{\bar{N}}_{N}\eta_{\bar{M}\bar{N}}$. These matrices obey the Clifford algebra $\{\Gamma^M, \Gamma^N\}=+2g^{MN}\mathds{1}_8$. Here, $D_M$ is the gauge covariant derivative given by \cite{Liu:2007gk, Liu:2007eb, Parameswaran:2006db}
\begin{eqnarray}\label{dcov}
D_M=\partial_M +\Omega_M  - iqA_M,
\end{eqnarray}
where $\Omega_M=\frac{1}{4}\eta^{\bar{P}\bar{N}}\xi^{\bar{M}}_{N}\left[\partial_{M}\xi^{N}_{\bar{P}}+\Gamma^{N}_{MQ}\xi^{Q}_{\bar{P}}\right]\Gamma_{\bar{M}}\Gamma_{\bar{N}}$ is the spin connection and $A_{M}=A_{\mu}(x)\hat{x}+A_{\theta}(r)\hat{\theta}$ is a cylindrically symmetric background  gauge vector field. We remark that here, $A_M$ is not related to the dynamical field $\mathcal{A}_{M}$ of the previous section.

The non-vanishing terms of the spin connection are
\begin{eqnarray}\label{spincon}
\Omega_{\mu}=\frac{1}{4}\frac{F^{\prime}(r)}{\sqrt{F(r)}}\Gamma_{\bar{\mu}}\Gamma_{\bar{r}}\quad \text{and} \quad \Omega_{\theta}=\frac{1}{4}\frac{H^{\prime}(r)}{\sqrt{H(r)}}\Gamma_{\bar{\theta}}\Gamma_{\bar{r}}.
\end{eqnarray}
Substituting the equations \eqref{dcov} and \eqref{spincon} in the action \eqref{action-fermions-6D} in the background metric given by Eq. \eqref{metric}, we obtain the Dirac equation:
\begin{equation}
\begin{split}
\label{Dirac-6D-2}
\Gamma^M D_M\Psi  = \Big[ & F^{-\frac{1}{2}}\Gamma^{\bar{\mu}}\Big(\partial_\mu-iqA_{\mu}(x)\Big) + \Gamma^{\bar{r}}\Big(\partial_r + \frac{F^{\prime}}{F} + \frac{H^{\prime}}{4H}\Big) + \\
                 & + H^{-\frac{1}{2}}\Gamma^{\bar{\theta}}\Big(\partial_{\theta}-iqA_{\theta}(r)\Big)\Big]\Psi=0.
\end{split}
\end{equation}

We will now choose the usual Weyl spinor and gamma matrices representation on six dimensional models \cite{Oda:2000zc, Liu:2007gk, Liu:2007eb, Parameswaran:2006db, Dantas:2013iha, Sousa:2014dpa}:
\begin{equation}
\label{rowspinor}
\Psi(x,r,\theta)=
\begin{pmatrix}
\psi_{4}\\
0\\
\end{pmatrix},
\end{equation}
\begin{eqnarray}
\label{Matrices-6D-4D}
\Gamma^{\bar{\mu}}=
\begin{pmatrix}
0 &\gamma^{\bar{\mu}}\\
\gamma^{\bar{\mu}} &0
\end{pmatrix},\quad
\Gamma^{\bar{r}}=
\begin{pmatrix}
0 &\gamma^{5}\\
\gamma^{5} &0
\end{pmatrix},\quad
\Gamma^{\bar{\theta}}=
\begin{pmatrix}
0 &-\gamma^{\theta}\\
\gamma^{\theta} &0\\
\end{pmatrix}.
\end{eqnarray}
Due to the metric sign convention $(-,+,+,+,+,+)$ the $\gamma^{\mu}$ matrices in Weyl representation become:
\begin{eqnarray}
\gamma^0= -i\sigma^{1}\otimes\mathds{1}_2, \quad \gamma^{i}=-\sigma^2\otimes\sigma^i, \quad \gamma^5=\sigma^3\otimes\mathds{1}_2, \quad \gamma^{\theta}=i\mathds{1}_4,
\end{eqnarray} 
where $\sigma^i$ are Pauli matrices and $\mathds{1}$ is the identity matrix. In this convention, the matrix $\gamma^0$ is anti-hermitian, while the others are hermitian. The $\gamma^5$ is such that $\gamma^{5}\psi_{R,L}=\pm \psi_{R,L}$. The Dirac operator acts as $\gamma^{\mu}\left(\partial_{\mu}-iqA_{\mu}\right)\psi=m\psi$. Others representations can be directly deduced from the general forms of the Ref. \cite{Budinich:2001nh} for signature $(+,-,-,-,-,-)$.

Now, let us perform a Kaluza-Klein decomposition on $\psi_{4}$ in the form
\begin{equation}
\label{spinorkkdecomposition}
\psi_4(x,r,\theta)=\frac{1}{\sqrt{2\pi}}\sum\limits_{n, l}\Big[\psi_{R_{n,l}}(x)\alpha_{R_{n,l}}(r)+\psi_{L_{n,l}}(x)\alpha_{L_{n,l}}(r)\Big]\e^{il\theta}.
\end{equation}

Using the equations \eqref{rowspinor}, \eqref{Matrices-6D-4D} and \eqref{spinorkkdecomposition} for the $s$-wave solution \cite{Oda:2000zc, Liu:2007gk, Liu:2007eb}, the Dirac equation (\ref{Dirac-6D-2}) turns to the following chiral coupled equations
\begin{eqnarray}\label{Dirac-6D-4}
\begin{cases}
\left[\partial_r +\mathcal{P}(r) + \mathcal{W}(r)\right]\alpha_{R_n}(r)=- \frac{m_n}{\sqrt{F(r)}}\alpha_{L_n}(r)\\
\left[\partial_r + \mathcal{P}(r) - \mathcal{W}(r)\right]\alpha_{L_n}(r)= \frac{m_n}{\sqrt{F(r)}}\alpha_{R_n}(r),
\end{cases}
\end{eqnarray}
where
\begin{equation}\label{p}
\mathcal{P}(r) =  \frac{F^{\prime}(r)}{F(r)}+\frac{H^{\prime}(r)}{4 H(r)} = -c\left[\frac{5}{4}\tanh^2{(c r)-2\sech^2{(cr)}\coth{(cr)}}\right]
\end{equation}
and
\begin{equation}\label{q}
\mathcal{W}(r) = q\frac{A_{\theta}(r)}{\sqrt{H(r)}} =  c q\frac{A_{\theta}(r)}{\tanh{(c r)}}\e^{\frac{1}{2}[cr-\tanh{(cr)}]}.
\end{equation}



\subsection{Spin $1/2$ Zero Mode}\label{Sec_ZeroMode}

For $m=0$, the expressions \eqref{Dirac-6D-4} decouple in two first order differential equations which solutions are:
\begin{eqnarray}\label{alphar0b}
\alpha^{0}_{R_n,L_s}(r) = C_{0} \exp{\left[-\int_{r^{\prime}}{\left(\mathcal{P} \pm \mathcal{W}\right)}dr^{\prime}\right]},
\end{eqnarray}
where $C_{0}$ is a normalization constant.

In order to the zero-mode be localized and free of singularities at the origin, we impose the orthogonality condition
\begin{eqnarray}\label{anorm} \int_{0}^{\infty}{\lvert\alpha_{R_n,L_s}(r)\lvert^2}dr^{\prime}=\delta_{R_n,L_s}.
\end{eqnarray}
This condition implies that $\lim_{r \to \infty}\alpha^0_{R,L}(r)=0$.
Nonetheless, since
\begin{equation}\label{ar}
-\int_{r^{\prime}} {\mathcal{P}(r)}dr^{\prime}=\frac{5}{4}\left[cr -\tanh{(cr)}+\frac{2}{5}\ln{\left(\frac{\tanh{(cr)}}{c}\right)}\right]
\end{equation}
is non-convergent, the $A_{\theta}(r)$ function presented in Eq. \eqref{q} has to be adjusted in order to overcome this drawback.
Assuming that
\begin{eqnarray}\label{qfix}
\mathcal{W}(r)=-\lambda \mathcal{P}(r),
\end{eqnarray}
where $\lambda$ is a dimensionless coupling constant, the \textit{rigth-handed} zero mode becomes
\begin{equation}
\alpha^0_{R}(r) = C_{0} \exp\left[ \int_{r^{\prime}}{dr^{\prime}}\left(\lambda-1\right)\mathcal{P} \right]=C_{0}  F^{(\lambda-1)}(r)H^{\frac{1}{4}(\lambda-1)}(r).
\end{equation}
For $\lambda=0$ this expression is the same obtained in Ref. \cite{Oda:2000zc} which is non-normalizable.

Using the explicit expressions of the warp factors in Eq. \eqref{Eq_WarpFactors} for string-cigar, the zero mode and the gauge  angular component are given, respectively, by
\begin{equation}\label{alphar0R}
\alpha^0_{R}(r) = C_{0}\left(\frac{\tanh{(cr)}}{c}\right)^{\frac{(\lambda-1)}{2}}\exp{\left(\frac{5}{4}(1-\lambda)[cr + \tanh{(cr)}]\right)}
\end{equation}
and
\begin{eqnarray}\label{athetar}
A_{\theta}(r)=\frac{\lambda}{q}\left[\frac{5}{4}\tanh^3{(c r)}-2\sech^2{(cr)}\right]\e^{-\frac{1}{2}[cr-\tanh{(cr)}]}.
\end{eqnarray}
In the absence of the coupling ($\lambda = 0$), the massless mode is not localized in the brane. For $\lambda > 1$, the zero mode is normalizable, but only for $\lambda > 3$ its derivative is continuous and null at the origin. From these restrictions over $\lambda$, we find the following boundary conditions
\begin{eqnarray}\label{cc}
\begin{cases}
\alpha^0_{R,L}(0)= \displaystyle\lim_{r \to \infty} \alpha^0_{R,L}(r)=0\\
\partial_r\left[\alpha^0_{R,L}(r)\right]_{r=0}=\displaystyle \lim_{r \to \infty} \partial_r\left[ \alpha^0_{R,L}(r)\right]= 0.
\end{cases}
\end{eqnarray}

\begin{figure}[!htb] 
       \begin{minipage}[t]{0.50 \linewidth}
           \includegraphics[width=\linewidth]{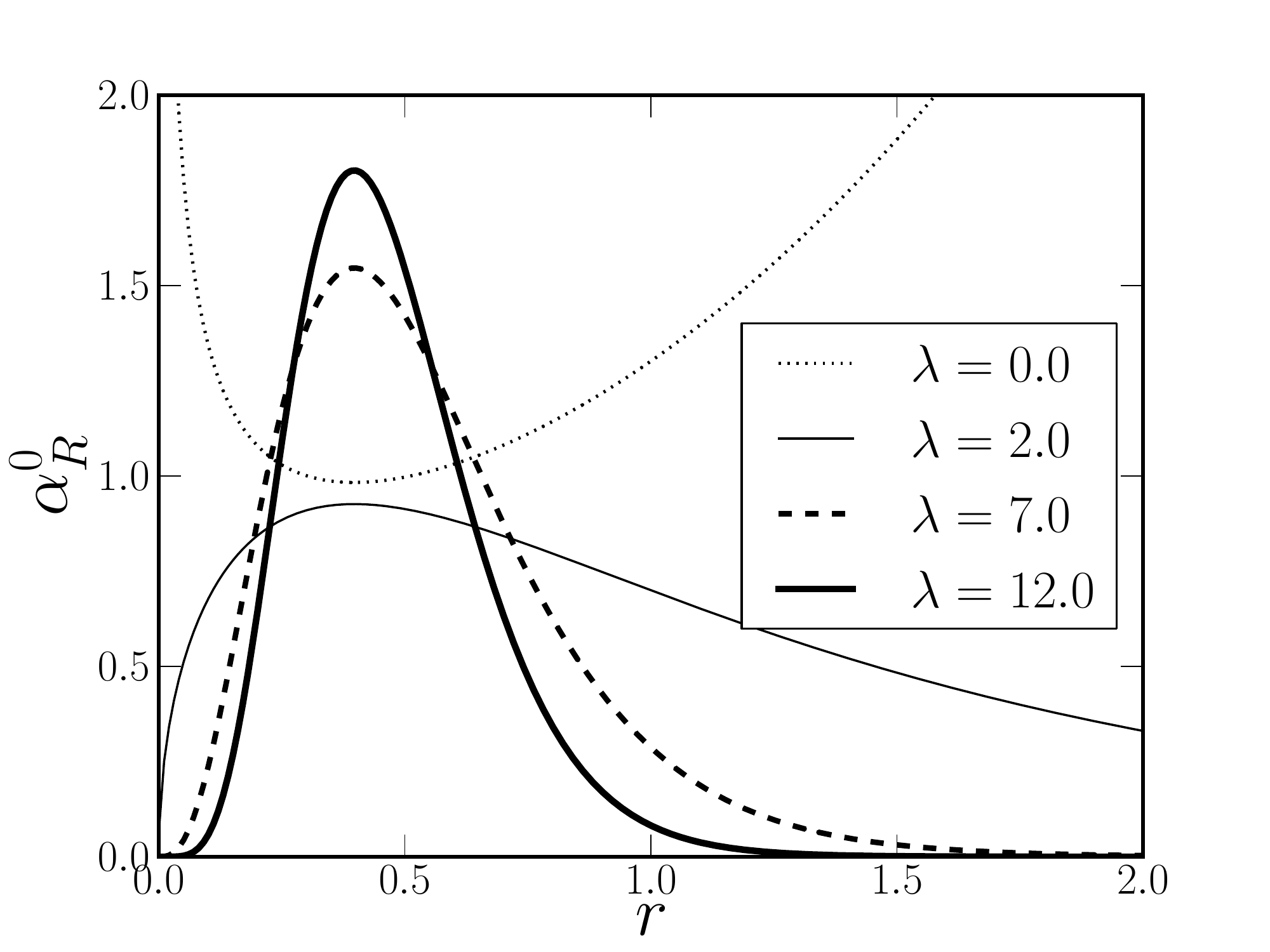}\\
           \caption{Plot of the right-handed fermionic zero mode for $c = 0.5$ in the String-Cigar model}
          \label{Fig_ZeroMode}
       \end{minipage}\hfill
       \begin{minipage}[t]{0.50 \linewidth}
           \includegraphics[width=\linewidth]{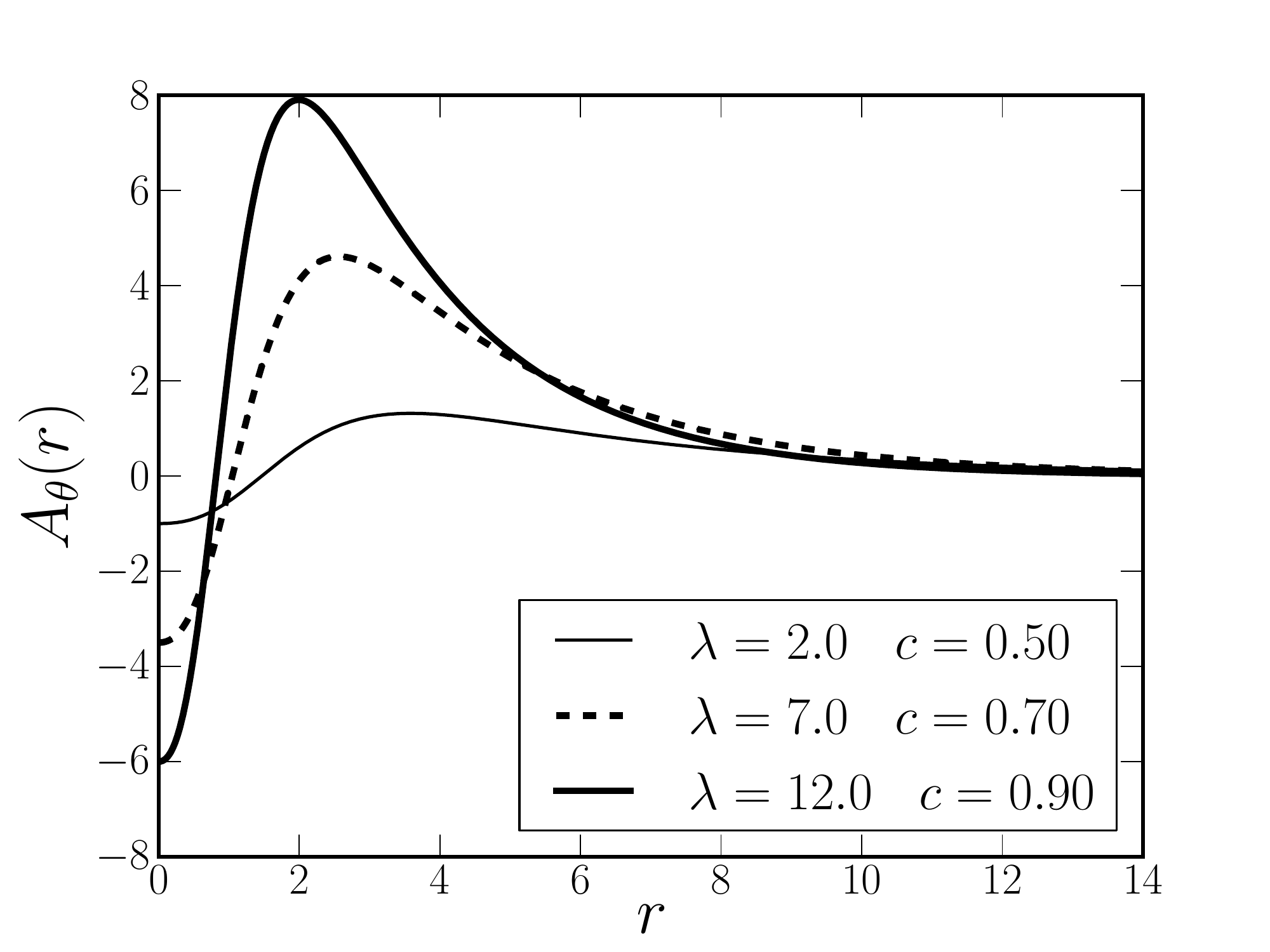}\\
           \caption{Plot of the gauge angular component for the String-Cigar model.}
           \label{Fig_Atheta}
       \end{minipage}
   \end{figure}

We plot the fermionic zero mode and the gauge angular term in the figures \ref{Fig_ZeroMode} and \ref{Fig_Atheta}, respectively. In both cases, the coupling constant $\lambda$ controls the amplitude, whereas the geometric parameter $c$ regulates their distribution over the radial extra-dimension. The displacement of the zero mode from the origin is an important result present in expression \eqref{alphar0R}. This feature is closely related to the fact that the brane core is not placed at $r = 0$ \cite{Silva:2012yj}. This zero mode has a similar profile to that of the energy density of the string-cigar model \cite{Silva:2012yj}. Note, in the figure \ref{Fig_ZeroMode}, that the zero mode satisfies the homogeneous boundary conditions \eqref{cc} provided that $\lambda>3$.


In order to confine the massless  \textit{left-handed}  fermions we would have to make $\lambda \to-\lambda$ in the Eq. \eqref{qfix} with the same restriction $\lvert \lambda\lvert >3$. Therefore, only one massless chiral mode can be trapped in the brane. This is a well-known result in five warped dimensional models \cite{CASA-Fermion-TwoField-ThickBrane, Chineses-Ressonance, Wilami1,  Correa:2010zg, German:2012rv, Castro:2010uj}.

It is interesting to note at this point that the references \cite{Oda:2000zc,  Liu:2007gk} use a less restrictive imposition on the radial component $\alpha^0_{R,L}(r)$ in the form
\begin{eqnarray}\label{anorm2}
I_{\frac{1}{2}}=\int_{0}^{\infty}{\hat{I}_{\frac{1}{2}}(r)}dr^{\prime}=\int_{0}^{\infty}{\sqrt{-g}F^{-\frac{1}{2}}(r)\lvert\alpha^0_{R_n,L_s}(r)\lvert^2dr^{\prime}}=\delta_{R_s,L_n},
\end{eqnarray}
which comes from the effective action $S^0_{eff}(x,r,\theta)$ \cite{Oda:2000zc, Liu:2007gk} using the equations \eqref{action-fermions-6D} and \eqref{spinorkkdecomposition}, namely
\begin{eqnarray}\label{ieff}
S^0_{eff}=\int_{-\infty}^{\infty}{\overline{\psi}(x)i\gamma^{\mu}\left[\partial_{\mu} + iqA_{\mu}(x)\right]\psi(x)}d^4x^{\prime} \int_0^{\infty}{ \hat{I}_{\frac{1}{2}}(r)}dr^{\prime}\int_0^{2\pi}{\frac{d\theta}{2\pi}}.
\end{eqnarray}
However, if we adopt only the condition above, the spinor in the string-cirgar scenario will be exposed to singularities, while for the condition \eqref{anorm}, these singularities vanish. Besides, the Eq. \eqref{anorm2} is satisfied too.


\subsection{Spin $1/2$ Massive Modes}\label{Sec_MassiveModes}

In order to study the massive modes, let us decouple the system of first-order differential equations \eqref{Dirac-6D-4} performing the conformal change of variable $z(r)$ given by Eq. \eqref{Eq_z(r)} which turns the coupled first-order differential equations system \eqref{Dirac-6D-4} to
\begin{eqnarray}\label{stlz}
\left(\partial^2_z+2\tilde{\mathcal{P}}\partial_z+\left[\tilde{\mathcal{P}}^2 -\tilde{\mathcal{W}}^2+ \left(\dot{\tilde{\mathcal{P}}} \pm \dot{\tilde{\mathcal{W}}}\right)\right]\right)\alpha_{R_n,L_n}(z)=
-m^2_n\alpha_{R_n,L_n}(z),
\end{eqnarray}
where
\begin{equation}\label{pqz}
\tilde{\mathcal{P}}=\mathcal{P}(z)\sqrt{F(z)}\quad \text{,} \quad \tilde{\mathcal{W}}=\mathcal{W}(z)\sqrt{F(z)}.
\end{equation}
Here, the over-dots mean derivatives with respect to $z$ and the $+,-$ sign stands for the right and the left chirality, respectively.

The decoupled system of two second-order equations (\ref{stlz}) is composed by independent Sturm-Liouville problems for each chirality. Thus, we can analyse the dynamics for the chiralities independently. However, due to the involved form of the metric components \eqref{Eq_WarpFactors}, the conformal coordinate $z(r)$
in Eq. \eqref{Eq_z(r)} can not be achieved analytically in general for the string-cigar geometry. The functions $\mathcal{P}(z)$ and $\mathcal{W}(z)$ (and their derivatives) must be constructed from a numerical integral of Eq. \eqref{Eq_z(r)}. In order to avoid the cumulative round-off errors in the forthcoming analysis, we will study the  Eq. \eqref{stlz} in the $r$ coordinate, where the metric functions are already defined. It turns out that returning to the $r$ coordinate, the second-order system \eqref{stlz} is still decoupled, and it can be written as
\begin{equation}
\begin{split}
\alpha^{\prime\prime}_{R_n,L_n}(r) + & \Big[ 3f + \frac{1}{2}g \Big]\alpha^{\prime}_{R_n,L_n}(r) + \Bigg\{  \frac{(1\mp\lambda)}{8}\Big[5f^2+fg +10f^{\prime} + 2g^{\prime}\Big] + \\
& + (1-\lambda^2)\Big[\frac{5}{4}f+\frac{g}{4}\Big]^2 \Bigg\}\alpha_{R_n,L_n} = -\frac{m_n^2}{F}\alpha_{R_n,L_n}(r)
\end{split}
\label{Eq_SturmLiouville}
\end{equation}


where
\begin{equation}
f(r) = \frac{F^{\prime}(r)}{F(r)} \hspace{0.5cm} \text{and} \hspace{0.5cm} g(r) = \frac{\beta^{\prime}(r)}{\beta(r)}.
\end{equation}

\subsubsection{Thin string}\label{spin1/2-massive-string-an}
For the thin string-like model ($f=-c$, and $g = 0$), the Sturm-Liouville KK equation \eqref{Eq_SturmLiouville} reduces to
\begin{equation}
\begin{split}
\alpha_{R_n,L_n}''(r) - & 3c\alpha_{R,L}'(r) + \frac{5c^{2}}{8}\Big[ (1\mp\lambda) + \frac{5}{2}(1 - \lambda^2)\Big]\alpha_{R_n,L_n}(r)=- m_n\e^{cr} \alpha_{R_n,L_n}(r) .
\end{split}
\label{thinstringslequation}
\end{equation}
For $\lambda=0$ (absence of the coupling), the Sturm-Liouville equation \eqref{thinstringslequation} turns to
\begin{equation}
\alpha_{R,L}''(r) -  3c\alpha_{R,L}'(r) + \frac{35}{16}c^2\alpha_{R,L}(r)=-m_n\e^{cr} \alpha_{R_n,L_n}(r),
\label{thinstringslfreeequation}
\end{equation}
and the solutions have the form
\begin{eqnarray}
\label{thinstringmassivemodefree}
\alpha_{R,L}=A^1_{R,L}\e^{\frac{3cr}{2}}\left[J_{\pm\frac{1}{2}}\left(\frac{2m}{c}\e^{\frac{cr}{2}}\right) + B^1_{R,L} Y_{\pm\frac{1}{2}}\left(\frac{2m}{c}\e^{\frac{cr}{2}}\right)\right],
\end{eqnarray}
where $A^1_{R,L}, B^1_{R,L}$ are integration constants. It is worthwhile to mention that
the massive modes depend on the Bessel functions  of order $\mu_{R,L}=\pm \frac{1}{2}$, while the graviton has order $5/2$ \eqref{GS-MassiveModes} \cite{Gherghetta:2000qi}
and the gauge vector field has order $3/2$ \eqref{chiequation} \cite{Oda:2000zc,Oda:2003jc}.

Moreover, unlike the graviton, gauge vector field and scalar field, the fermionic massless mode for $\lambda=0$ in not localized on the thin string-like brane \cite{Oda:2000zc}. In fact, for $m=0$ in \eqref{thinstringslfreeequation}, the massless solution has the form
%
\begin{equation}
\label{thinstringmasslessmodefree}
\alpha^0_{R,L}=A^0_{R,L}\e^{p_1 cr} + B^0_{R,L}\e^{p_2 cr},
\end{equation}
where $p_{1,2}=\frac{12\pm \sqrt{109}}{8}$ and $A^0_{R,L}, B^0_{R,L}$ are integration constants. 

\begin{figure}[htb]
 \centering
    \includegraphics[width=0.50\textwidth]{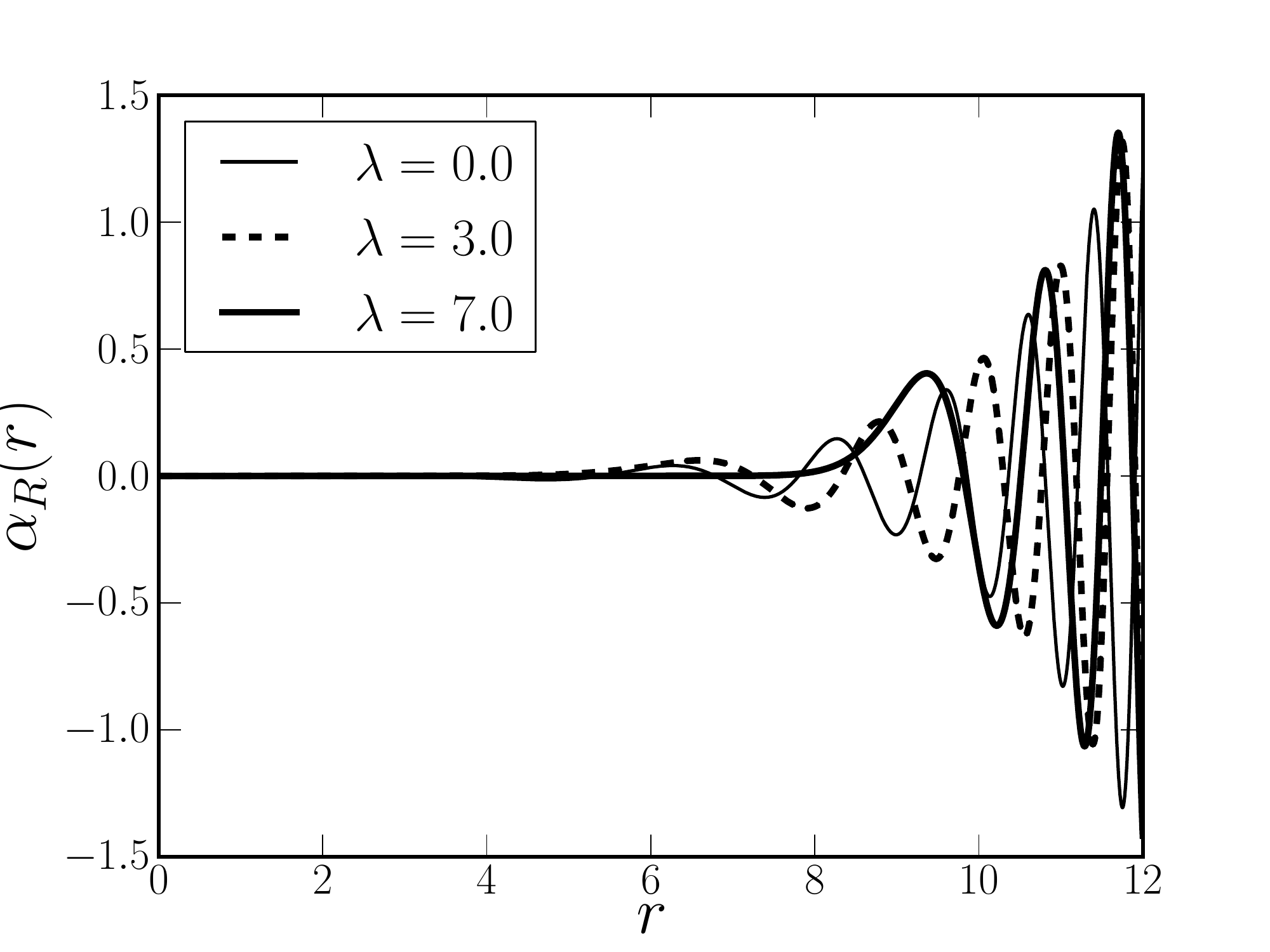}
 \caption{Massive mode in the thin string-like (GS) scenario for $m = 0.50$. The values of the parameters was $A_R = B_R = 1.5 \times 10^{-3}$, $\lambda = 5.0$ and $c = 0.5$.}
 \label{Fig_Autovetor-Corda}
\end{figure}

For $\lambda\neq 0$, we write the solution as
\begin{eqnarray}
\label{thinstringmassivemode}
\alpha_{R,L}=A_{R,L}\e^{\frac{3cr}{2}}\left[J_{\mu_{R,L}}\left(\frac{2m}{c}\e^{\frac{cr}{2}}\right) + B_{R,L} Y_{\mu_{R,L}}\left(\frac{2m}{c}\e^{\frac{cr}{2}}\right)\right],
\end{eqnarray}
where $A_{R,L}, B_{R,L}$ are integration constants and
\begin{equation}
\mu_{R,L}=(5\lambda\pm 1)/2
\end{equation}
are the orders of the Bessel functions. The massive modes in Eq. \eqref{thinstringmassivemode} bears a resemblance with those found in 5D for massive fermions \cite{gp}, where the mass, as the gauge coupling $\lambda$ here, controls the order of the Bessel functions. We plot in the figure \ref{Fig_Autovetor-Corda} the analytical right handed solution \eqref{thinstringmassivemode} for different values of $\lambda$. Note that the gauge coupling distances the massive modes from the brane.

Unlike the massless mode, the massive eigenfunctions \eqref{thinstringmassivemode} are not trapped in the thin string-like brane due to the exponential and the Bessel functions.  Nevertheless, in order to satisfy the boundary conditions, the order of the Bessel function ought to be $\mu_{R}>7$ and $\mu_{L}>8$. The coupling allows the Bessel functions order to be integer or half-integer. For $\lambda$ even the order $\mu_{R,L}$ is half integer whereas for $\lambda$ odd
$\mu_{R,L}$ is an integer. Although the coupling constant $\lambda$ can be any real number, the Bessel functions of irrational order suffers of branch issues and then, henceforward, we shall be concerned with the rational $\lambda$ only.
A noteworthy feature is that the massive modes are related by $\mu_{R}=\mu_{L}+1$. An interesting reason for this symmetry will be shown in the next section through the Schr\"{o}dinger approach.

Applying the boundary conditions \eqref{cc} in the massive modes \eqref{thinstringmassivemode} at the origin and at some cutoff distance $r=r_{\mx}$, for $m\ll c$
we obtain the conditions $B_{R,L}=0$ and
\begin{equation}
\label{spectrumbesselfunction}
J_{\mu_{R,L}}\left(\frac{2m}{c}\e^{\frac{cr_{\mx}}{2}}\right)=0.
\end{equation}
From the roots of the Bessel function \eqref{spectrumbesselfunction},
we find that the KK massive spectrum $m_{n}$ is discrete  and it behaves as the series \citep{watson}
\begin{equation}
\label{thingstringmassivespectrum}
m_{n}\approx \frac{c\pi}{2}\e^{-\frac{cr_{max}}{2}}\left[n+\frac{2\mu_{R,L}-3}{4} + \frac{\mu_{R,L}}{2}\frac{(2-\mu_{R,L})}{\left(n+\frac{2\mu_{R,L}-3}{4}\right)\pi^{2}} + \mathcal{O}\left(\frac{1}{n^2}\right) \right].
\end{equation}
The KK spectrum \eqref{thingstringmassivespectrum} exhibits an increasing behaviour, as expected \cite{Oda:2000zc}. For large $n$, the spectrum behaves linearly whereas for small $n$ the $\mathcal{O}\left(\frac{1}{n}\right)$ terms in the series \eqref{thingstringmassivespectrum} changes the rate of increasing of the masses. The mass gap between the massless and the first massive mode is given by
\begin{equation}
\Delta=m_{0}\approx \frac{c\pi}{2}\e^{-\frac{cr_{max}}{2}},
\end{equation}
which vanishes for an infinite radial coordinate. Then, for an infinite radial extra dimensions, there is no mass gap between the massless mode and the massive KK tower, as usual in warped compactified models \citep{rs,Gherghetta:2000qi,Oda:2000zc}.

\subsubsection{String-cigar model}
For the string-cigar model, where
\begin{equation}
f(r) = -c\tanh^2{(cr)} \hspace{0.5cm} \text{and} \hspace{0.5cm} g(r) = 2c\frac{\sech{(cr)}}{\tanh{(cr)}},
\end{equation}
the eigenvalue problem \eqref{Eq_SturmLiouville} is quite complex to be studied analytically. Then, we employ numerical methods to find the mass spectrum $m_n$ and the correspondent eigenfunctions. The numerical integration of the Eq. $(\ref{Eq_SturmLiouville})$ was performed using the matrix method \cite{MatrixMethod} based on finite differences with second order truncation error. To avoid the singularity in $r = 0$ and the overflow errors provided by the exponential functions for large $r$, we discretized the domain $r \in [0.01, 13.00]$ into an uniform grid with constant stepsize $h = 0.01$.

\begin{figure}[htb]
 \centering
    \includegraphics[width=0.50\textwidth]{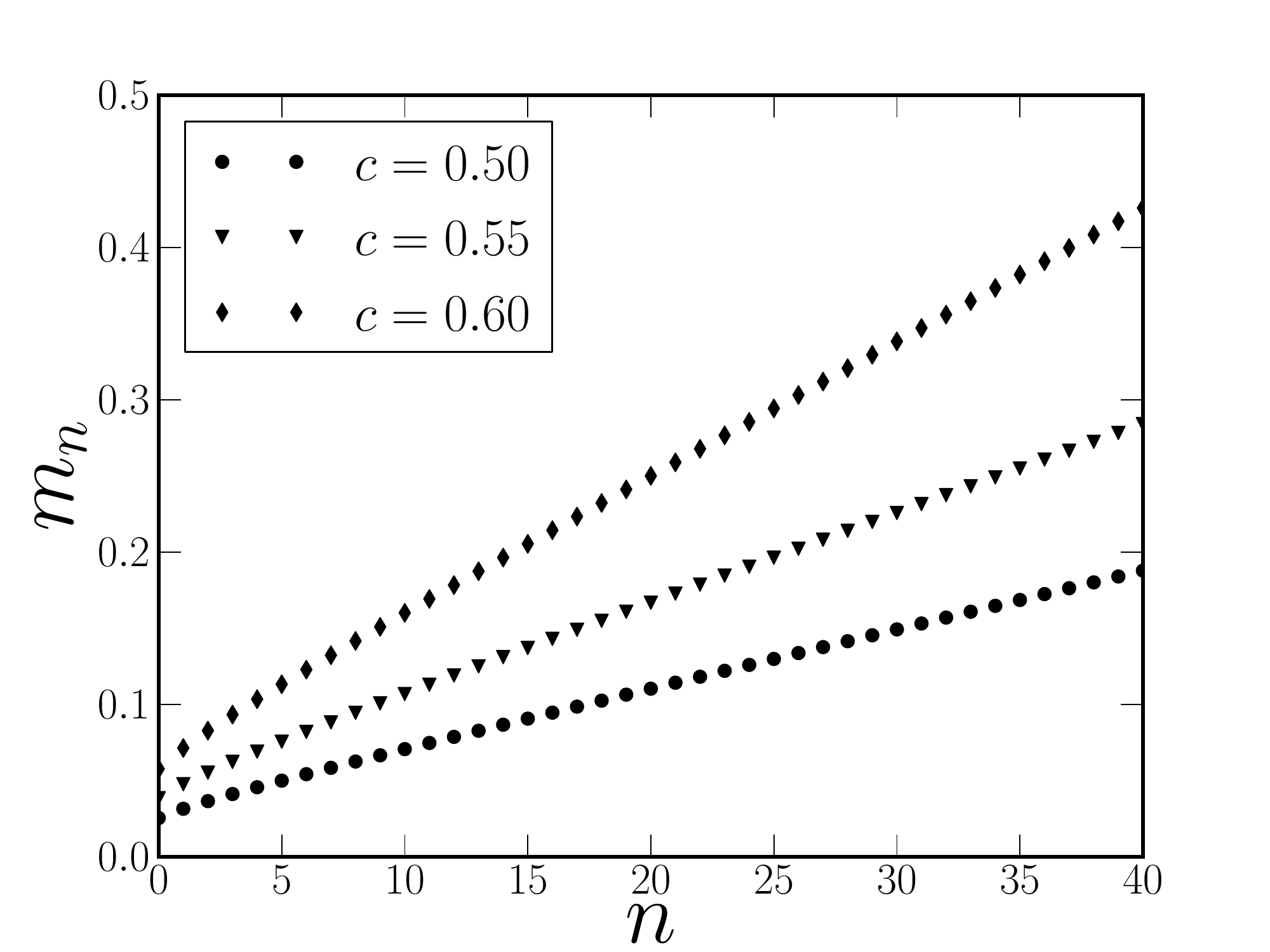}
 \caption{Mass spectrum for $\lambda = 7.0$ in the String-Cigar model. Note that heavier masses will be acceptable when the bulk Planck mass increases. Moreover, the growth rate is slightly lower for the first indexes.}
 \label{Fig_Spectra}
\end{figure}

We plot the lowest mass eigenvalues in the figure $\ref{Fig_Spectra}$ for $\lambda = 7.0$ and different values of the geometric parameter $c$, which is related to the bulk Planck mass \cite{Silva:2012yj}. Note that the spectrum is monotonically increasing, as expected from the Kaluza-Klein theories. This regime is valid for $m \ll c$ \cite{Oda-PRD}. Further, heavier masses will be acceptable as $c$ increases. Moreover, the growth rate of $m_n$ is slightly lower for the first indexes $n$. This is in accordance with Eq. \eqref{thingstringmassivespectrum}. It is worthwhile to mention that, although we treat both chiralities left and right in Eq. $\eqref{Eq_SturmLiouville}$ in an independent way, the relation among the eingenstates reveals that, regardless the massless mode, the left and right spectrum are the same.

\begin{figure}[!htb] 
\begin{minipage}[t]{0.45 \linewidth}
                \includegraphics[width=\linewidth]{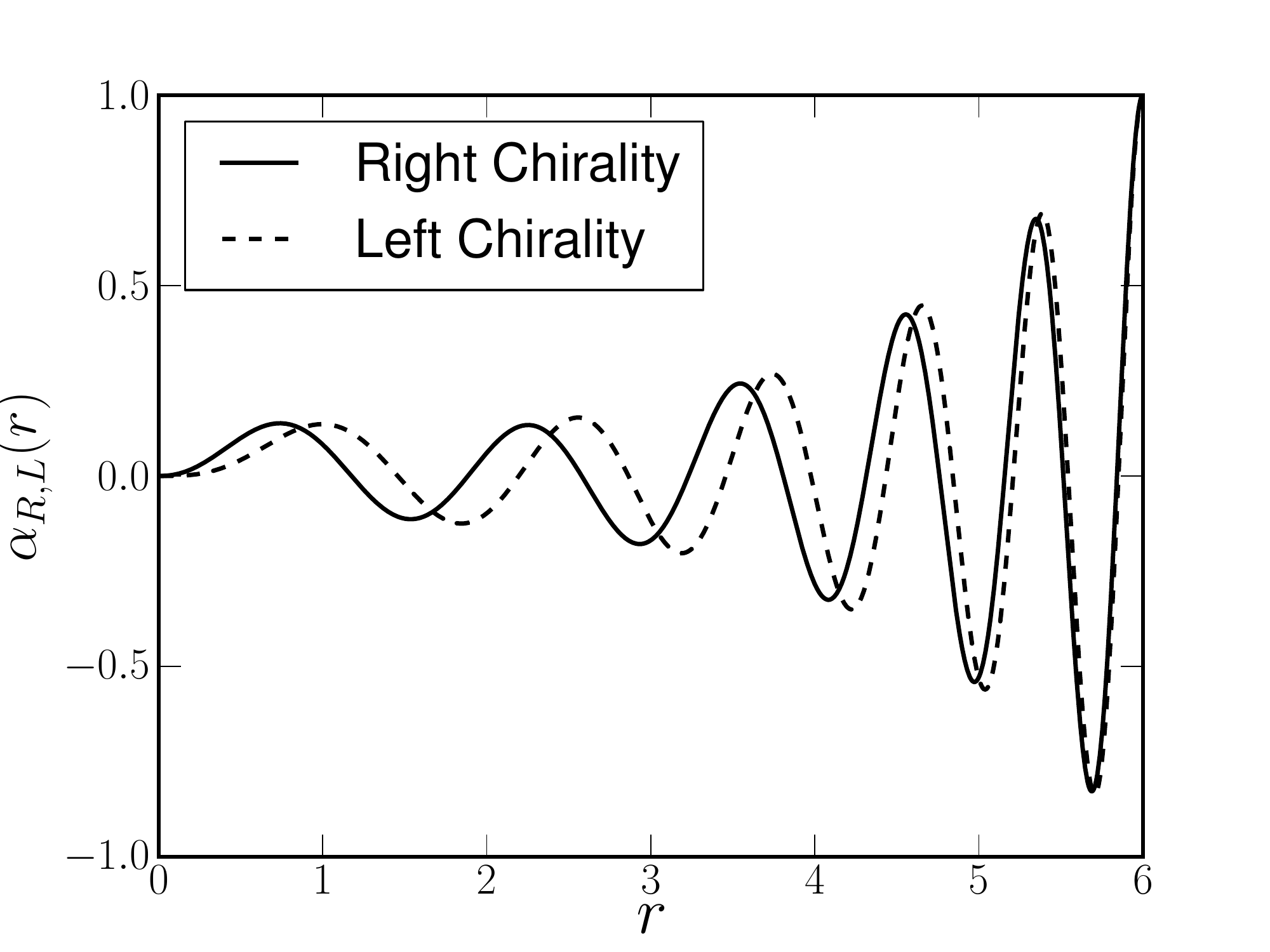}
                \caption{Normalized eigenfunctions for $c = 0.5$ and $\lambda = 5.0$ in the String-Cigar model. The masses eigenvalues was obtained as  $m_R = 0.4024$ and $m_L = 0.4025$.}
                \label{Fig_Autovetor1}
\end{minipage}
~,\qquad
        ~ 
\begin{minipage}[t]{0.45 \linewidth}
                \includegraphics[width=\linewidth]{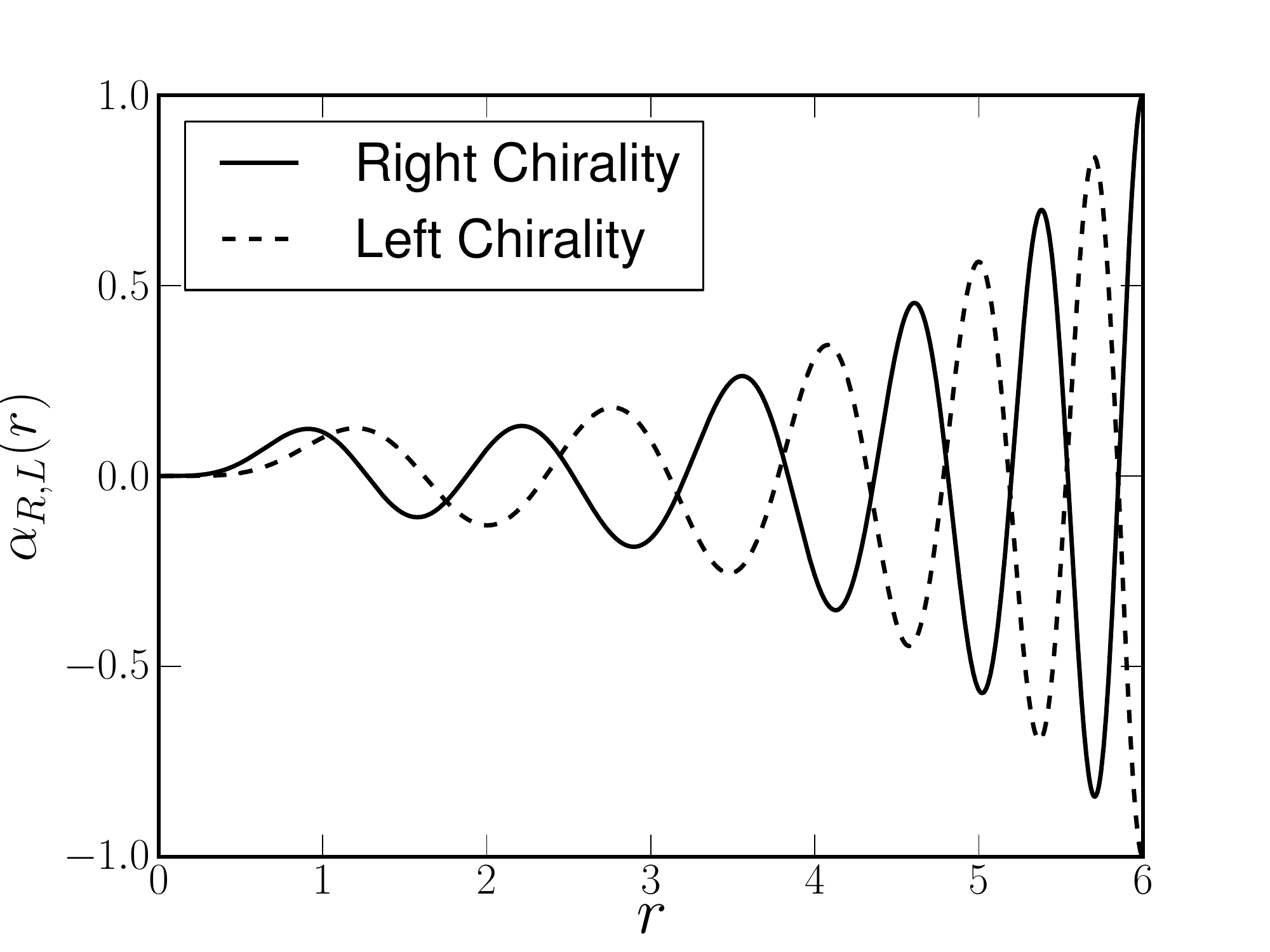}
                \caption{Normalized eigenfunctions for $c = 0.5$ and $\lambda=9.0$ in the String-Cigar model. The masses eigenvalues was obtained as  $m_R = 0.4593$ and $m_L = 0.4586$. }
                \label{Fig_Autovetor2}

        ~ 

\end{minipage}
\end{figure}

In the Figures $\ref{Fig_Autovetor1}$ and $\ref{Fig_Autovetor2}$, we present the eigenfunctions for $c = 0.5$ and for $\lambda = 5.0$ and $9.0$, respectively. Near the brane, they behave as Bessel functions of integer $(> 2)$ order which increases with $\lambda$. Since the string-cigar model recovers the thin-string one asymptotically, it is expected that the eigenfunctions have the same behaviour for large $r$ \cite{Silva:2012yj}. Note that this occurs when compared with Figure \ref{Fig_Autovetor-Corda}. Moreover, the core of the string-cigar brane amplify the massive modes near the origin. This behaviour occurs for the gravitational \cite{Graviton_Charuto} and gauge \cite{Costa:2015dva} fields. This is the first stimulus for searching resonant modes. In the next section, we will present the formalism concerning resonant states.


\subsection{Schr\"{o}dinger approach}
\label{spin1/2schrodinger}
The equality of right and left spectra and the relation between the eingenfunctions result from an underlying symmetry which is manifested in the Schr\"{o}dinger approach. Indeed, performing the change of variable
\begin{equation}
\alpha_{R,L}(z)=\exp \left[-\int_{z^{\prime}}{\tilde{\mathcal{P}}(z)}dz^{\prime}\right]\tilde{\alpha}_{R,L}(z),
\label{Eq_SecondTransformation}
\end{equation}
in the Eq. \eqref{stlz}, we obtain a Schr\"odinger-like equation as
\begin{eqnarray}
\label{Eq_Schroedinger}
\left[-\partial_z^2 + V_{R,L}(z)\right]\tilde{\alpha}_{R,L}(z)=m^2\tilde{\alpha}_{R,L}(z),
\end{eqnarray}
where
\begin{equation}
\label{Eq_Potential}
V_{R,L}(z)=\tilde{\mathcal{W}}^2(z)\pm \partial_z\tilde{\mathcal{W}}.
\end{equation}

For the thin string-like brane, the analogue potential has the form
\begin{equation}
V_{R,L}(z)=\frac{5\lambda}{2}\left[\frac{5\lambda}{2}\mp 1\right]\frac{1}{\left(z+\frac{2}{c}\right)^2}.
\end{equation}
For $\lambda > 0.4$, there is a potential barrier at the origin for both chiralities. Since the potential
vanishes asymptotically, there is no mass gap for an infinite radial coordinate, as we found using the Sturm-Liouville
approach in section \ref{Sec_MassiveModes}.
Defining the variable
\begin{equation}
x:=m\left(z+\frac{2}{c}\right),
\end{equation}
the Schrodinger-like equation for the thin string-like model reads
\begin{eqnarray}
\label{Eq_Schroedinger2}
\left[-\partial_x^2 + \Bigg\{\frac{5\lambda}{2}\Big[\frac{5\lambda}{2}\mp 1\Big]\frac{1}{x^2}\Bigg\}-1\right]\tilde{\alpha}_{R,L}(x)=0,
\end{eqnarray}
whose solution can be written as
\begin{equation}
\label{thinstringschrodingermassivemode}
\tilde{\alpha}_{R,L}(x)=N_{R,L}\sqrt{x}\Big[J_{\mu_{R,L}}(x)+A_{R,L}Y_{\mu_{R,L}}(x)\Big].
\end{equation}
The Eq. \eqref{thinstringschrodingermassivemode} is only another form of the massive KK state given by Eq. \eqref{thinstringmassivemode}.

The structure of the potential \eqref{Eq_Potential} enable us to rewrite the Schr\"odinger-like equation as the system of
equations
\begin{eqnarray}
\label{rightschrodingerequation}
H_{R}\tilde{\alpha}_{R}(z)= m^2\tilde{\alpha}_{R}(z),\quad H_{L}\tilde{\alpha}_{L}(z)	=	m^2\tilde{\alpha}_{L}(z),
\label{leftschrodingerequation}
\end{eqnarray}

%
where the Hamiltonians operator $H_{R,L}$ can be factorized into
\begin{eqnarray}
\label{hamiltonianfactorization}
H_{R}	=		A^{\dagger}A, \quad H_{L}	=	AA^{\dagger}
\end{eqnarray}

and
\begin{equation}
\label{schrodingerfirstorderoperator}
A(z):=\frac{d}{dz}+\mathcal{\tilde{W}}(z).
\end{equation}
For the thin string-like model, the first-order differential operator $A(z)$ writes
\begin{equation}
A(z)=\frac{d}{dz} - 5\frac{\lambda}{2}\frac{1}{\left(z+\frac{2}{c}\right)}.
\end{equation}

The analogue Hamiltonian operators in Eq.s \eqref{rightschrodingerequation} and \eqref{leftschrodingerequation} form an analogue Supersymmetric Quantum Mechanics structure. In fact, defining the analogue charge operators \cite{susy}
\begin{eqnarray}
Q=\left(\begin{array}{cc}
0 & 0 \\
	  A & 0	
\end{array}\right)	&	,	& Q^{\dagger}=\left(\begin{array}{cc}
0 & A^{\dagger} \\
	  0 & 0	
\end{array}\right)
\end{eqnarray}
which are nilpotent, i.e., $Q^{2}=Q^{\dagger 2}=0$, and also defining the SUSY-like Hamiltonian \cite{susy}
\begin{equation}
H=\left(\begin{array}{cc}
H_{R}	&	0\\
0		&	H_{L}
\end{array}
\right)
\end{equation}
we obtain the SUSY-like quantum mechanics algebra \cite{susy}
\begin{eqnarray}
H	&	=	&	\{Q,Q^{\dagger}\},
\end{eqnarray}
and $[Q,H]=[Q,H]=0$. The Hamiltonians $H_{R,L}$ are related by $H_{L}^{\dagger}=H_{R}$ and are called Hamiltonian superpartners whereas $\mathcal{\tilde{W}}$ is known as the superpotential \cite{CASA-Fermion-TwoField-ThickBrane,susy}.

One remarkable feature of the SUSY-like system \eqref{hamiltonianfactorization} is that the massive KK spectrum is the same for the both chiralities \cite{Csaki1,Csaki2,CASA-Fermion-TwoField-ThickBrane,CASA-Fermion-TwoField-ThickBrane}. Indeed, consider a massive eigenfunction $\tilde{\alpha}_{R}(z)$ of the right-handed Hamiltonian $H_{R}$ \eqref{rightschrodingerequation} with mass $m_{R}$. Define the function
\begin{equation}
\label{rightleftrelationsymmetry}
\tilde{\alpha}_{L}:=\frac{1}{m_{R}}A\tilde{\alpha}_{R}.
\end{equation}
Applying the left-handed Hamiltonian $H_{L}$ on the function $\tilde{\alpha}_{L}$, we find that $H_{L}\tilde{\alpha}_{L}(z)=m_{R}^2\tilde{\alpha}_{L}(z)$,
i.e., $\tilde{\alpha}_{L}$ is a left-handed eigenfunction with the same mass of $\tilde{\alpha}_{R}$. Defining \cite{susy}
\begin{equation}
\label{leftrightrelationsymmetry}
\tilde{\alpha}_{R}:=\frac{1}{m_{L}}A^{\dagger}\tilde{\alpha}_{L},
\end{equation}
we have $H_{R}\tilde{\alpha}_{R}(z)=m_{L}^2\tilde{\alpha}_{R}(z)$,
i.e., $\tilde{\alpha}_{R}$ is a right-handed eigenfunction with the same mass of $\tilde{\alpha}_{L}$.
Therefore, for each right-handed eigenfunction $\tilde{\alpha}_{R}$ exists a left-handed eigenfunction $\tilde{\alpha}_{L}$ with the same mass and vice-versa.

The SUSY-like structure of the Hamiltonians $H_{R,L}$ also guarantees that the spectrum is bounded from below. In fact, multiplying any of the Hamiltonians $H_{R,L}$ by the dual eigenfunction $\tilde{\alpha}_{R,L}$, respectively, we obtain $||A\tilde{\alpha}_{R,L}||^2 = m^{2}||\tilde{\alpha}_{R,L}||^2$, and hence, $m\geq 0$. The absence of tachyonic (negative norm) KK modes guarantees the stability of the spectrum. Further, it also enables us to employ a probabilistic approach to find the resonant modes, as we will discuss in the next section.

The Hamiltonian factorization and the absence of negative norm states allow us to reduce the problem to find the ground state from a second-order differential equation to a first-order differential equation. In fact, for the right-handed massless mode $H_{R}\tilde{\alpha}_{R}^{0}=0 \Rightarrow ||A\tilde{\alpha}_{R}^{0}||=0$, and thereby,
$A\tilde{\alpha}_{R}^0=0$, whereas for the left-handed massless mode $\tilde{\alpha}_{L}^0$, $H_{L}\tilde{\alpha}_{L}^{0}=0 \Rightarrow A^{\dagger}\tilde{\alpha}_{L}^0=0$. Thus, the massless modes $\tilde{\alpha}_{R,L}^0$ satisfy the equation
\begin{equation}
\dot{\tilde{\alpha}}_{R,L}^0 \pm \mathcal{\tilde{W}}(z)\tilde{\alpha}_{R,L}^0=0,
\end{equation}
whose solution is given by
\begin{equation}
\label{groundstate}
\tilde{\alpha}_{R,L}^0=\e^{\mp \int_{0}^{z}{\mathcal{\tilde{W}}(z')dz'}},
\end{equation}
By Eq. \eqref{groundstate}, only one chiral massless mode is normalizable, i.e.,
localizable. Using the change of dependent variable Eq. \eqref{Eq_SecondTransformation}, we obtain the expression \eqref{alphar0b} for the massles mode.
Then, for $\lambda>0$, only the right-handed massless mode is localized on the brane.

\subsection{Resonant modes}\label{Sec_Ress}
In spite of the Kaluza-Klein massive modes are not localized at the brane, some massive states can exhibit a relatively large amplitude near the brane \cite{CASA-Fermion-TwoField-ThickBrane}. These states, known as resonant modes, can be obtained by the quantum mechanical analog structure of the massive modes \cite{Csaki1}.
The resonant modes occur for potentials that exhibit a potential well near the brane and for masses $m^2$ up to the maximum value of the potential barrier \cite{Csaki1,Csaki2}.

\begin{figure}[!htb] 
  \begin{minipage}[t]{0.45\linewidth}
                \includegraphics[width=\linewidth]{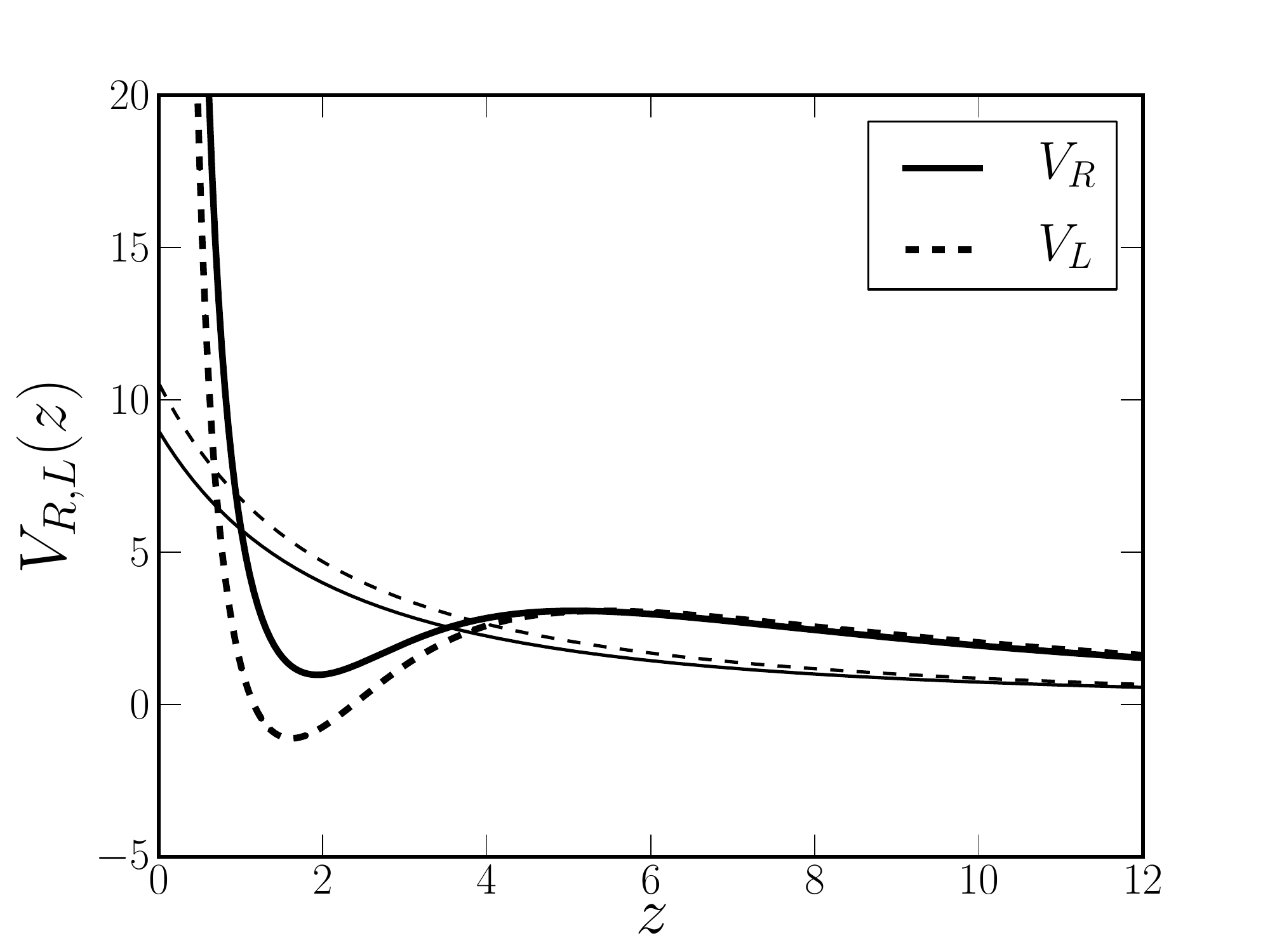}
                \caption{Potential function for both chirality for $c = 0.5$ and $\lambda = 5.0$. The thick lines correspond to the string-cigar model, while the thin lines, to GS one.}
                \label{Fig_Potential1}
  \end{minipage}%
        ~ 
  \begin{minipage}[t]{0.45\linewidth}
                \includegraphics[width=\linewidth]{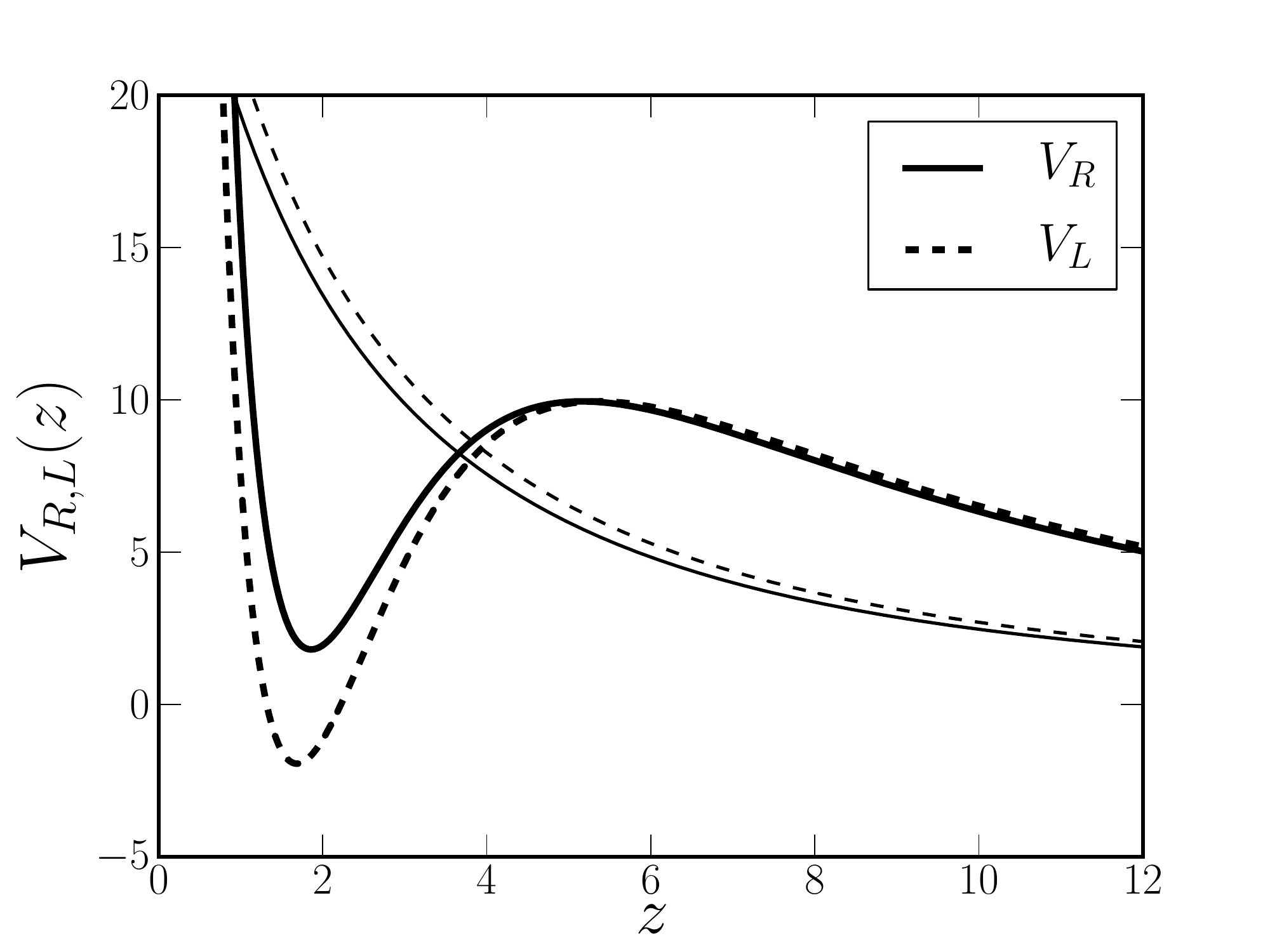}
                \caption{Potential function for both chirality for $c = 0.5$ and $\lambda = 9.0$. The thick lines correspond to the string-cigar model, while the thin lines, to GS one.}
                \label{Fig_Potential2}

        ~ 
         \end{minipage}
\end{figure}

In order to solve the Schr\"oedinger-like equation \eqref{Eq_Schroedinger2}, we need to construct the potential function $(\ref{Eq_Potential})$. For this, we calculated the warp factors and the gauge angular ansatz in the $z$-variable from the numerical integral of Eq. (\ref{Eq_z(r)}) using spline interpolation. We plot in the figures $\ref{Fig_Potential1}$ and $\ref{Fig_Potential2}$ the potential functions for both chiralities. Note that there is a potential well allowing the existence of bound states. The potential has the usual volcano shape. To find solutions of the Schr\"odinger-like equation \eqref{Eq_Schroedinger2} with the highest amplitudes near the brane (in comparison with its values far from the the defect), we used the resonance method \cite{CASA-Fermion-TwoField-ThickBrane}. The relative probability $P_{R,L}(m)$ to find a particle with mass $m$ in a narrow range $2\epsilon$ around the position $\bar{z}$ of the minimum of the potential well may be defined as \cite{CASA-Fermion-TwoField-ThickBrane, Chineses-Ressonance}
\begin{equation}
P_{R,L}(m) = \dfrac{1}{ \int_{z_{\mn}}^{z_{\mx}} |\tilde{\alpha}_{R,L}(z)|^2 dz} \int_{\bar{z} - \epsilon}^{\bar{z} + \epsilon} |\tilde{\alpha}_{R,L}(z)|^2 dz,
\label{Eq_Probabilidade}
\end{equation}
where $z_{\mn}$ and $z_{\mx}$ stand to the domain limits. To perform calculations near the minimum of the potential well, we adjusted $\epsilon = 0.1$.


\begin{figure}[!htb] 
\begin{minipage}[t]{0.45 \linewidth}

                \includegraphics[width=\textwidth]{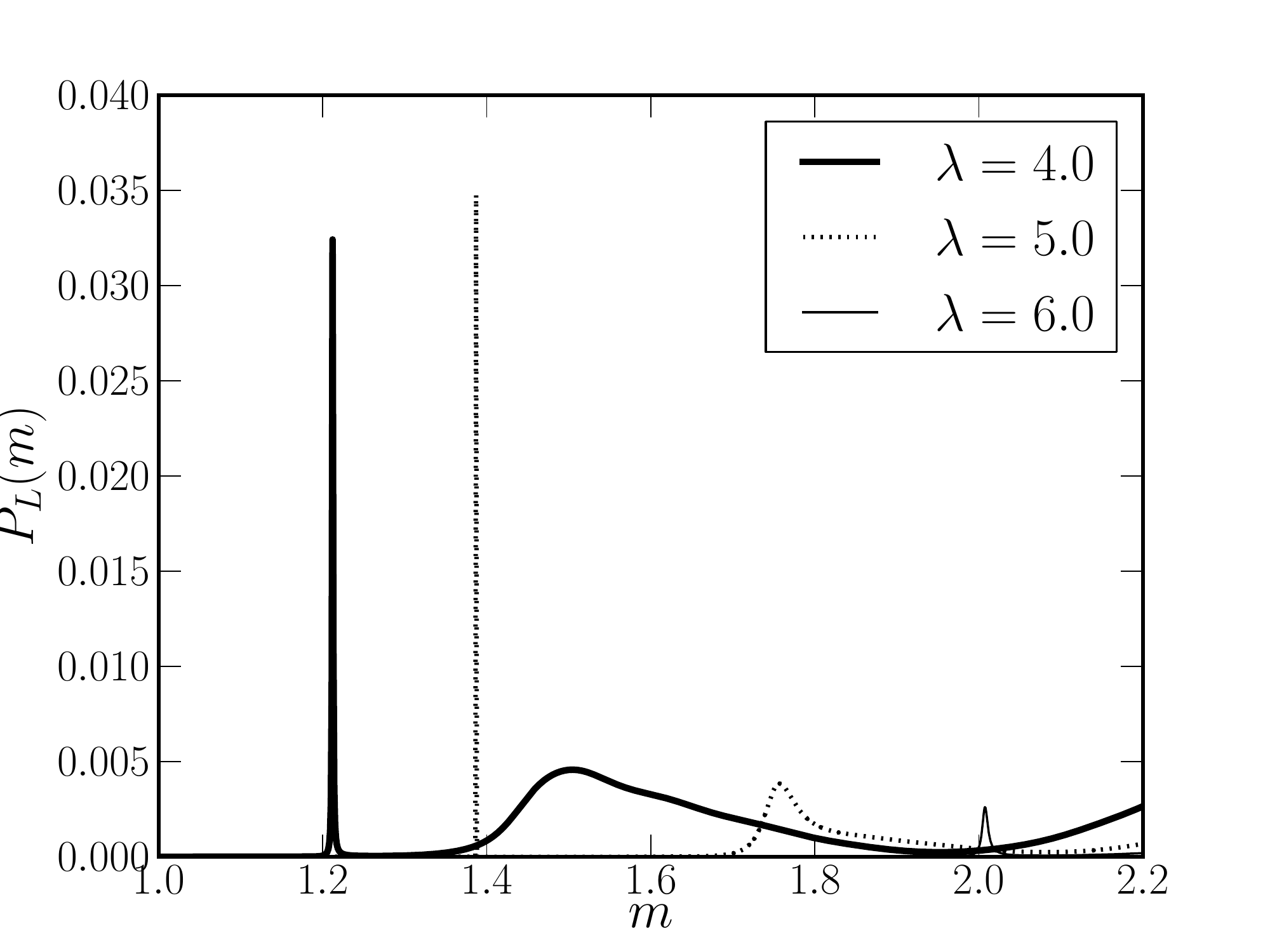}
                \caption{Plot of the probability distribution $P_{R,L}(m)$ for $c = 0.5$ (Left chirality) in the String-Cigar model.}
                \label{Fig_Resonance1}
   \end{minipage}%
        ~ 
    \begin{minipage}[t]{0.45\textwidth}
                \includegraphics[width=\textwidth]{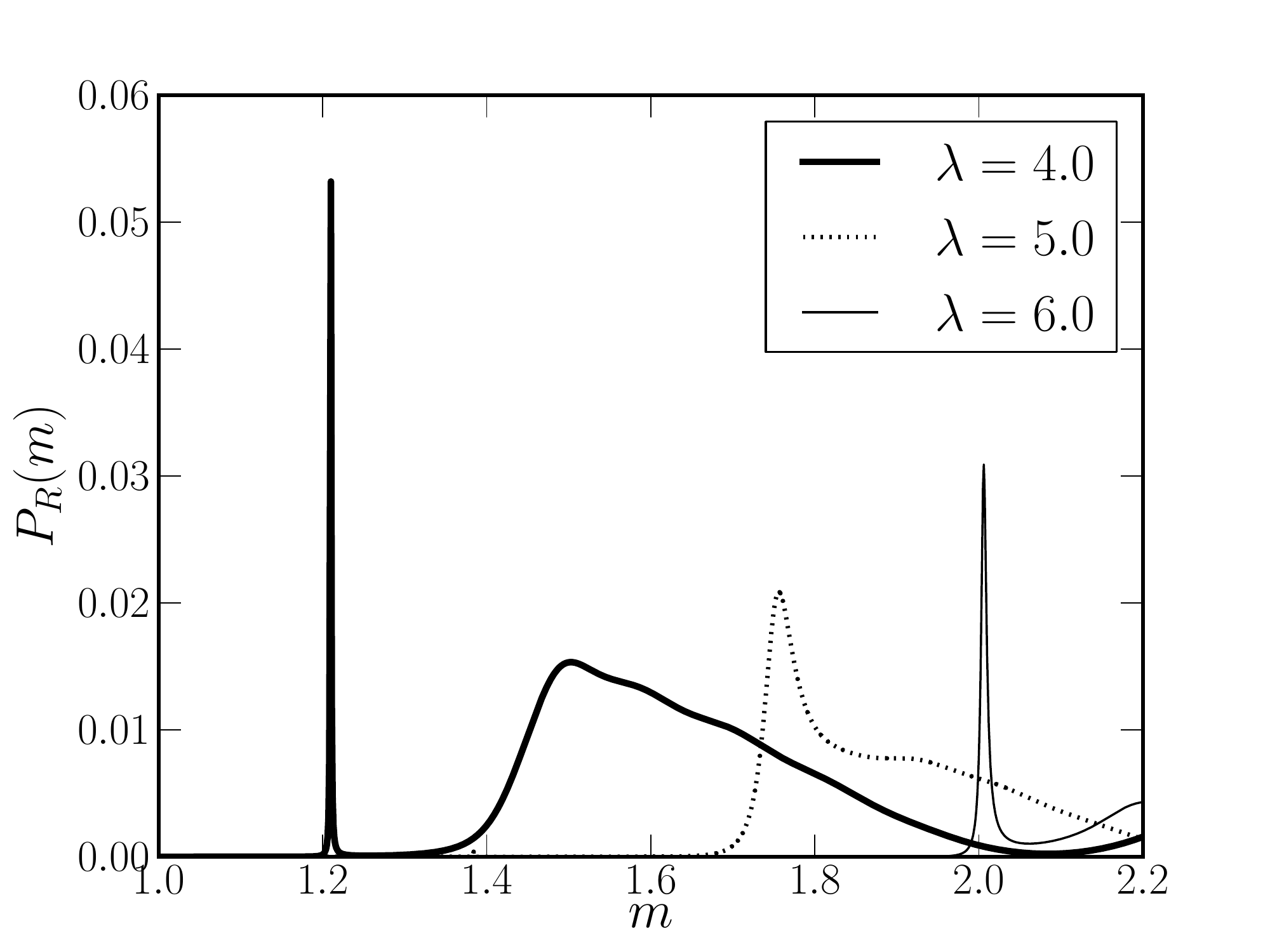}
                \caption{Plot of the probability distribution $P_{R,L}(m)$ for $c = 0.5$ (Right chirality) in the String-Cigar model.}
                \label{Fig_Resonance2}
        \end{minipage}
        ~ 

\end{figure}

\begin{figure}[!htb] 
\begin{minipage}[t]{0.45 \linewidth}

                \includegraphics[width=\textwidth]{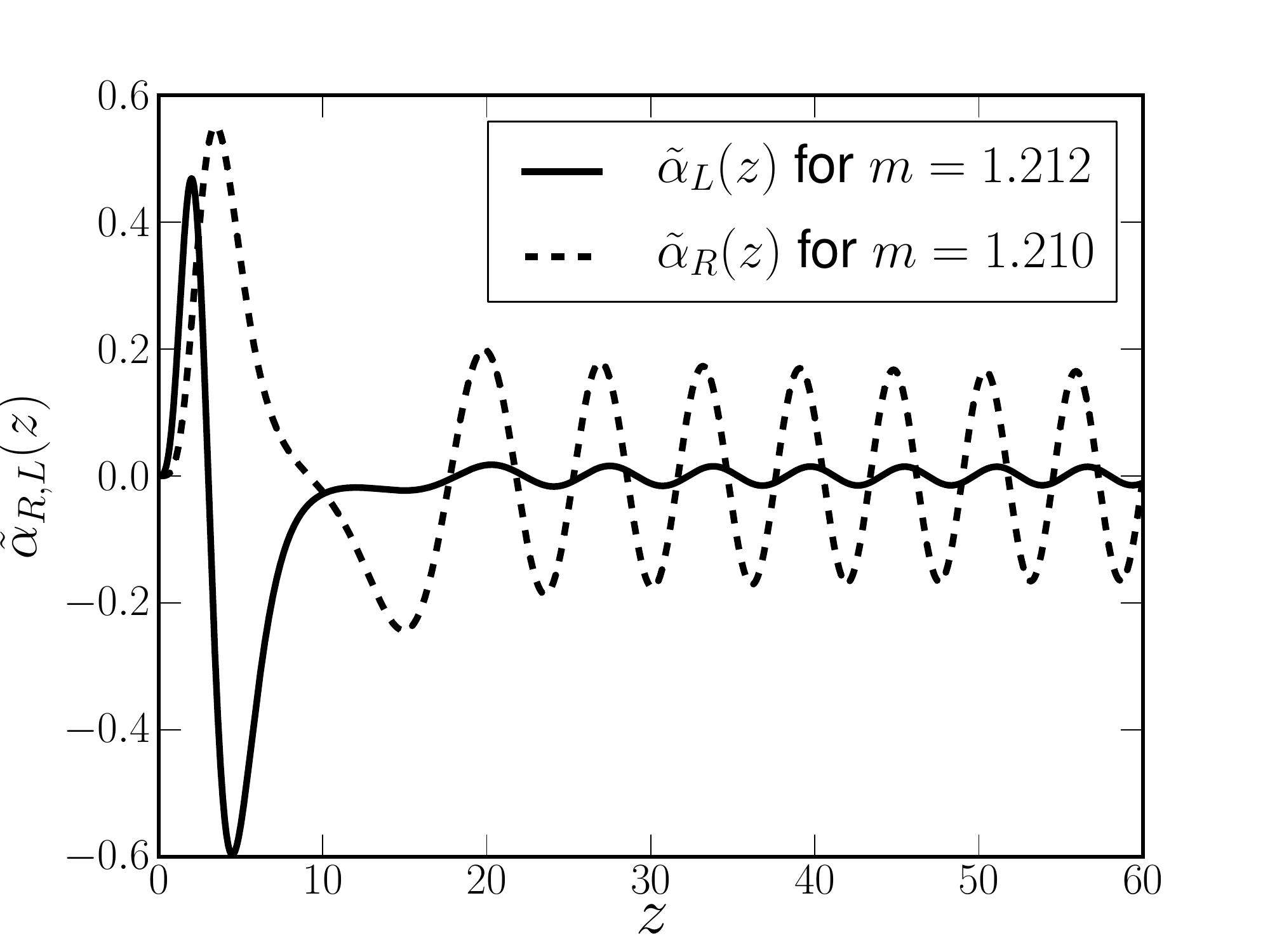}
                \caption{Normalized solutions of the Schr\"oedinger-like equationin the String-Cigar model for the masses corresponding to the peaks in the probability distribution. Left and right-handed solutions for $\lambda = 4.0$.}
                \label{Fig_FuncaoDeOnda_4}
        \end{minipage}%
        ~,\qquad
        ~ 
  \begin{minipage}[t]{0.45\textwidth}
                \includegraphics[width=\textwidth]{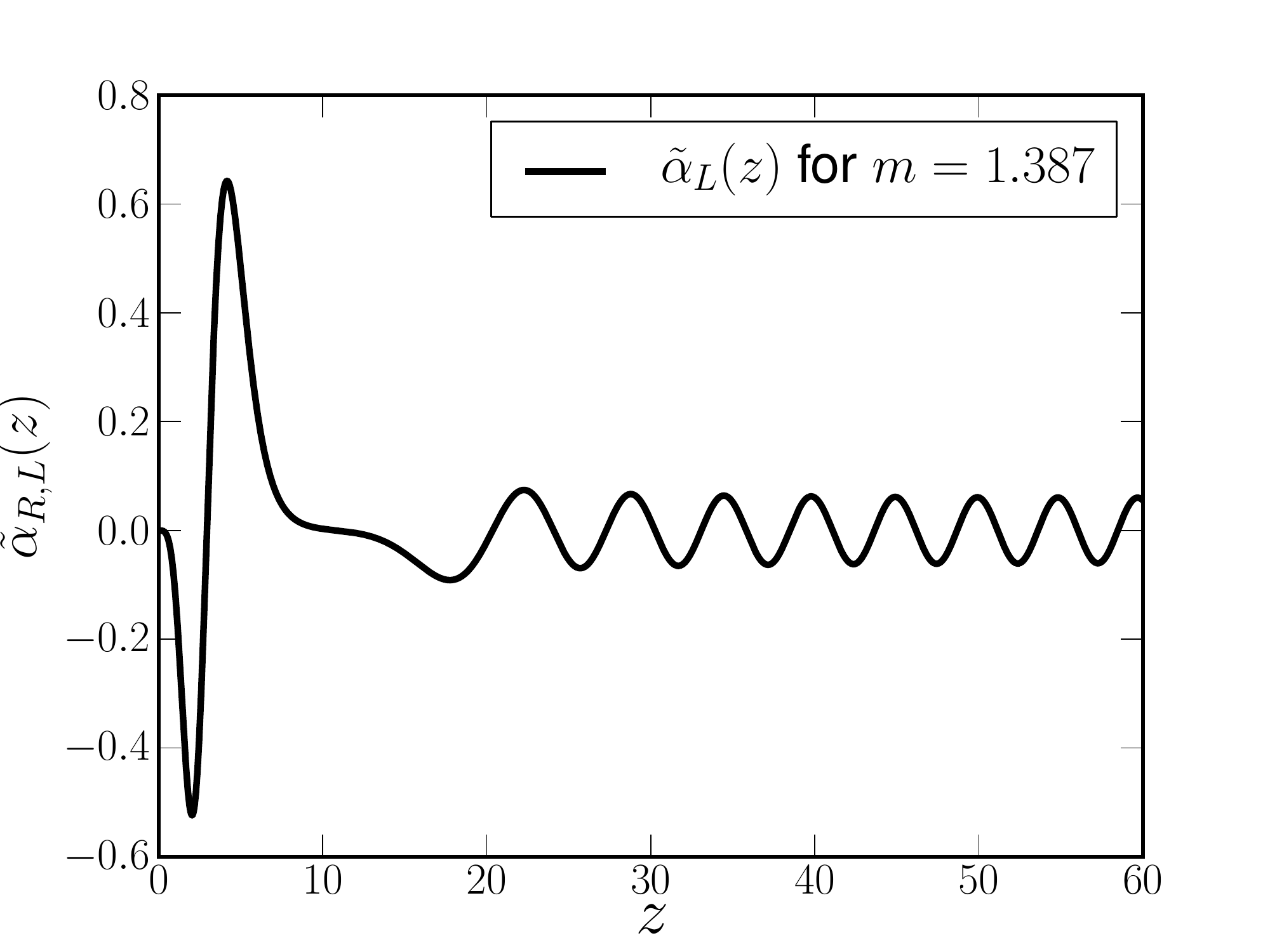}
                \caption{Normalized solution of the Schr\"oedinger-like equation in the String-Cigar model for the mass corresponding to the second peak in the probability distribution in the Fig. \ref{Fig_Resonance1}. This is the left-handed solution for $\lambda = 5.0$ and $m = 1.387$.}
                \label{Fig_FuncaoDeOnda_5}
        \end{minipage}
        ~ 

\end{figure}

We carried out the numerical integration of the Schr\"odinger-like equation $(\ref{Eq_Schroedinger2})$ for the potential function $(\ref{Eq_Potential})$ into Eq. \eqref{Eq_Probabilidade} using the Numerov algorithm \cite{Numerov} for a large sample of the parameters $c$ and $\lambda$. The distribution $P_{R,L}(m)$ exhibited peaks that may be referred as resonant modes \cite{CASA-Fermion-TwoField-ThickBrane}. In the figures \ref{Fig_Resonance1} and \ref{Fig_Resonance2}, we plot the function $P_{R,L}(m)$ for $c = 0.5$. For the left-handed case, there are very sharp peaks when $\lambda = 4.0$ and $\lambda = 5.0$, while for the right-handed one, when $\lambda = 4.0$ and $\lambda = 6.0$. However, only the first peak in $P_L$ represents a resonance. To verify this, we solved the Schr\"odinger-like equation for the masses corresponding to each peak in the distribution $P_{R,L}(m)$. The wave functions are plotted in figures \ref{Fig_FuncaoDeOnda_4} and \ref{Fig_FuncaoDeOnda_5}. Note that the solution $\tilde{\alpha}_L(z)$ for $\lambda = 4.0$ has the smallest oscillation far from the brane, which characterizes a resonant mode \cite{CASA-Fermion-TwoField-ThickBrane}. Although the peaks in the distributions $P_R(m)$ and $P_L(m)$ occur for masses very close (for $\lambda = 4.0$ when $c = 0.5$), only the left-handed case has a resonant feature.

Similar results were obtained for other values of $c$. In general, the coupling constant $\lambda$ determines the existence of a resonant mode, while the geometric parameter $c$ controls the ``position'' of the resonant peaks (i.e. the resonant mass). This is a expected result for a fixed $\lambda$, since varying the geometrical parameter $c$, which corresponds to change the Planck scale cut-off, different masses will be accepted as a resonant state. In a five dimensional thick brane scenario, the resonance of fermionic modes was also studied in Ref. \cite{Wilami1}.


\section{Spin $3/2$ fermionic field}
\label{Sec_Fermion-3/2}

In this section, we perform the confinement of the Rarita-Schwinger bulk field (spin $3/2$). Firstly, we start from the following action \cite{Oda:2000zc,Liu:2007gk}:
\begin{eqnarray}
\label{action-fermions-6D-3/2}
 S_{6_{3/2}}=\int{\sqrt{-g} \, \bar{\Psi}_{M} i\Gamma^{\left[M\right.}\Gamma^{N}\Gamma^{\left. P\right]} D_N \Psi_{P}}d^6x,
\end{eqnarray}
where square brackets denotes the anti-symmetrization. Its equation of motion has the form
\begin{eqnarray}\label{eqm3/2}
\Gamma^{\left[M\right.}\Gamma^{N}\Gamma^{\left. P\right]} D_N \Psi_{P}=0.
\end{eqnarray}
From now on, we will use the shorter notation $\Gamma^{MNP}$ to denote the product of matrices.

For this spin $3/2$ field, the covariant derivative gains an  additional term of affine connection when  compared to spin $1/2$ field \eqref{dcov}, namely
\begin{eqnarray}\label{dcov3/2}
D_M\Psi_N=\left(\partial_M +\Omega_M  - iqA_M\right)\Psi_N - \Gamma^{P}_{MN}\Psi_{P}.
\end{eqnarray}

The non-vanishing terms of the Eq. \eqref{dcov3/2} with the gauge imposition $\Psi_{\theta}=\Psi_r=0$ \cite{Liu:2007gk} are:
\begin{eqnarray}\label{di}
D_{\mu}\Psi_{\nu}=\left(\partial_{\mu}+\frac{1}{4}\frac{F^{\prime}}{F}\Gamma_{\mu}\Gamma_{r}-iqA_{\mu}\right)\Psi_{\nu},\\
D_{\mu}\Psi_{r}=-\frac{1}{2}\frac{F^{\prime}}{F}\Psi_{\mu},\quad
D_{r}\Psi_{\mu}=\left(\partial_{r}-\frac{1}{2}\frac{F^{\prime}}{F}\right)\Psi_{\mu},\label{dii}\\
D_{\theta}\Psi_{\mu}=\left(\partial_{\theta}+\frac{1}{4}\frac{H^{\prime}}{H}\Gamma_{\theta}\Gamma_{r}-iqA_{\theta}\right)\Psi_{\mu}.\label{diii}
\end{eqnarray}

%
%
%
%
%
%

Similarly to the decomposition of spin $1/2$ in Eq. \eqref{rowspinor}, the Refs. \cite{Oda:2000zc,Liu:2007gk, Liu:2007eb} exhibit the 4D Rarita-Schwinger vector-spinor in the form
\begin{equation}
\label{rowspinor3/2}
\Psi_{\mu}(x,r,\theta)=
\begin{pmatrix}
\psi_{\mu}^{(4)}\\
0\\
\end{pmatrix}
\end{equation}
Here the spinor $3/2$ assumes the  KK decomposition:
\begin{equation}
\label{spinorkkdecomposition3/2}
\psi_{\mu}^{(4)}(x,r,\theta)=\frac{1}{\sqrt{2\pi}}\sum\limits_{n, l}\Big[\psi_{\mu R_{n,l}}(x)u_{R_{n,l}}(r)+\psi_{\mu L_{n,l}}(x)u_{L_{n,l}}(r)\Big]\e^{il\theta},
\end{equation}
where the $4D$ section is  constrainted by $\partial^{\mu}\psi_{\mu}=\gamma^{\mu}\psi_{\mu}=0$  \cite{Oda:2000zc,Liu:2007gk} and  $\gamma^{\mu \nu \rho}\left(\partial_{\nu}-iqA_{\nu}\right)\psi_{\rho R,L }=m\gamma^{\mu\nu}\psi_{\nu L, R}$ \cite{Rahman:2011ik}.

Then, with these restrictions, the non-vanished terms of $\Gamma^{MNP}D_{N}\Psi_{P}$ are
\begin{eqnarray}\label{DDspin3/2}
\Gamma^{\rho\mu\nu}D_{\mu}\Psi_{\nu}=\Gamma^{\rho\mu\nu}\left(\partial_{\mu}-iqA_{\mu}\right)\Psi_{\nu}+\frac{F^{\prime}}{F}\Gamma^{ \nu \rho}\Gamma_{r}\Psi_{\nu},\\
\Gamma^{\rho \mu r}D_{\mu}\Psi_{r}=-\frac{1}{2}\frac{F^{\prime}}{F}\Gamma^{\rho \mu r}\Psi_{\mu},\quad \Gamma^{\rho r\mu}D_{r}\Psi_{\mu}=\Gamma^{\rho r\mu}\left(\partial_{r}-\frac{1}{2}\frac{F^{\prime}}{F}\right)\Psi_{\mu},\\
\Gamma^{\rho \theta \mu}D_{\theta}\Psi_{\mu}=\Gamma^{\rho\theta\mu}\left(\partial_{\theta}-iqA_{\theta}\right)\Psi_{\mu}+\frac{H^{\prime}}{4H}\Gamma^{\mu \rho}\Gamma_{r}\Psi_{\mu}.\label{DDspin3/2b}
\end{eqnarray}

Thus, writing the gamma matrices in the flat form ($\Gamma^{M}=\xi^{M}_{\bar{M}}\Gamma^{\bar{M}}$) and dropping down some indexes, for $l=0$, the equation of motion \eqref{eqm3/2} with equations \eqref{DDspin3/2}-\eqref{DDspin3/2b} becomes

\begin{eqnarray}
\label{zero3/2}
\begin{cases}
\left[\partial_r +\left(\mathcal{P}(r)-\frac{F^{\prime}}{2F}\right) + \mathcal{W}(r)\right]u_{R_n}(r)=- \frac{m_n}{\sqrt{F(r)}}u_{L_n}(r)\\
\left[\partial_r + \left(\mathcal{P}(r)-\frac{F^{\prime}}{2F}\right) - \mathcal{W}(r)\right]u_{L_n}(r)= \frac{m_n}{\sqrt{F(r)}}u_{R_n}(r),
\end{cases}
\end{eqnarray}
with $\mathcal{P}(r)$ defined in Eq. \eqref{p} and $\mathcal{W}(r)$ in \eqref{q}. We conclude that the equation \eqref{zero3/2} is similar to the spin $1/2$ case presented in the equation \eqref{Dirac-6D-4} with the additional term $-\frac{F^{\prime}}{2F}$.

\begin{figure}[!htb] 
\begin{minipage}[t]{0.45 \linewidth}
\includegraphics[width=0.99\textwidth]{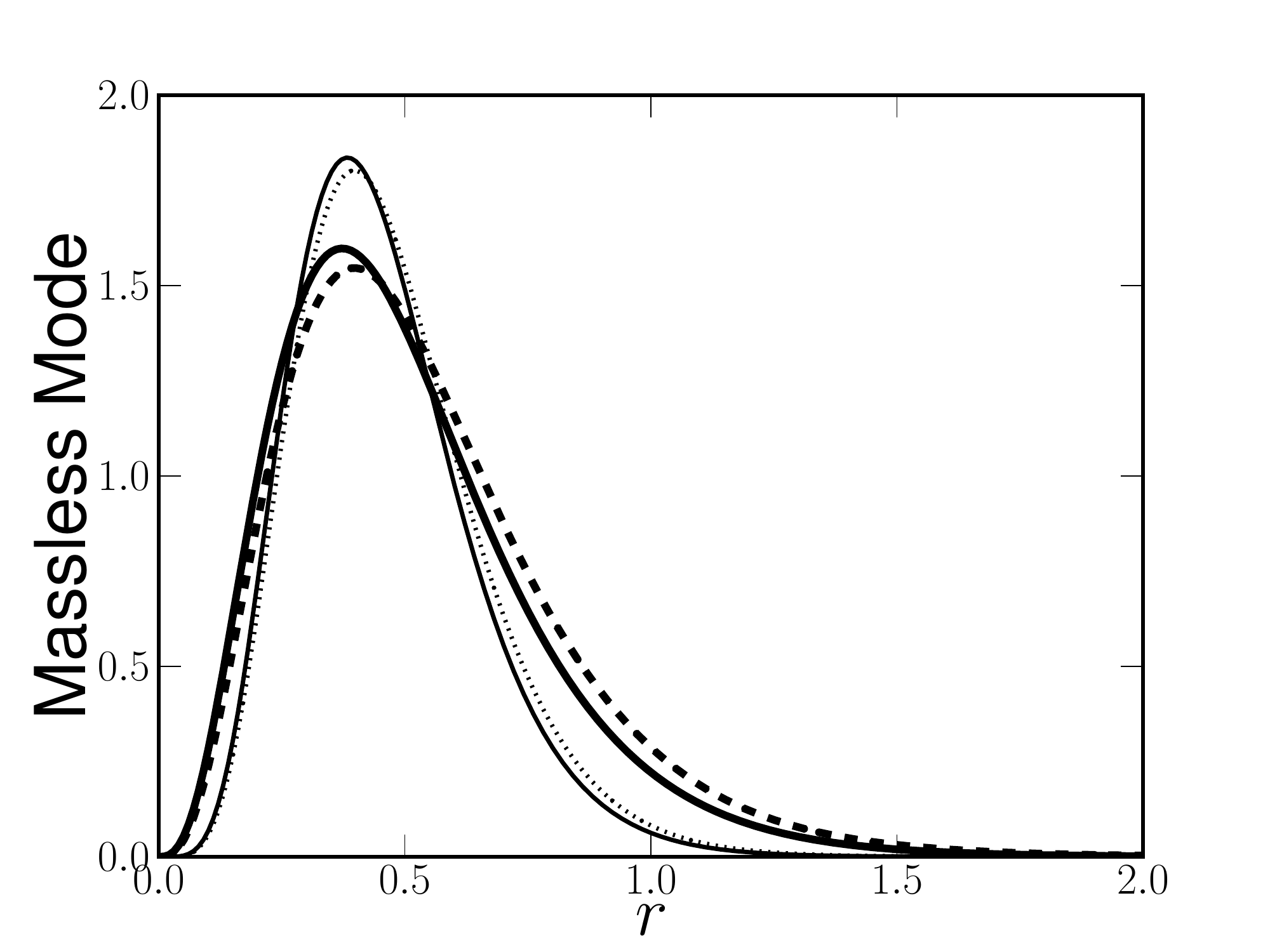}
 \caption{Comparative plot of right handed zero mode of spinorial fields in the String-Cigar model.  Filled lines corresponds to Rarita-Schwinger field, while dashed ones, to spin 1/2 field. The values for lambda was set to 7 (broader) and 12 (sharper). The geometric parameter was set $c=0.5$ in both cases.}
 \label{Fig_MasslessMode-Rarita}
        \end{minipage}%
        ~,\qquad
        ~ 
  \begin{minipage}[t]{0.45\textwidth}
                \includegraphics[width=0.99\textwidth]{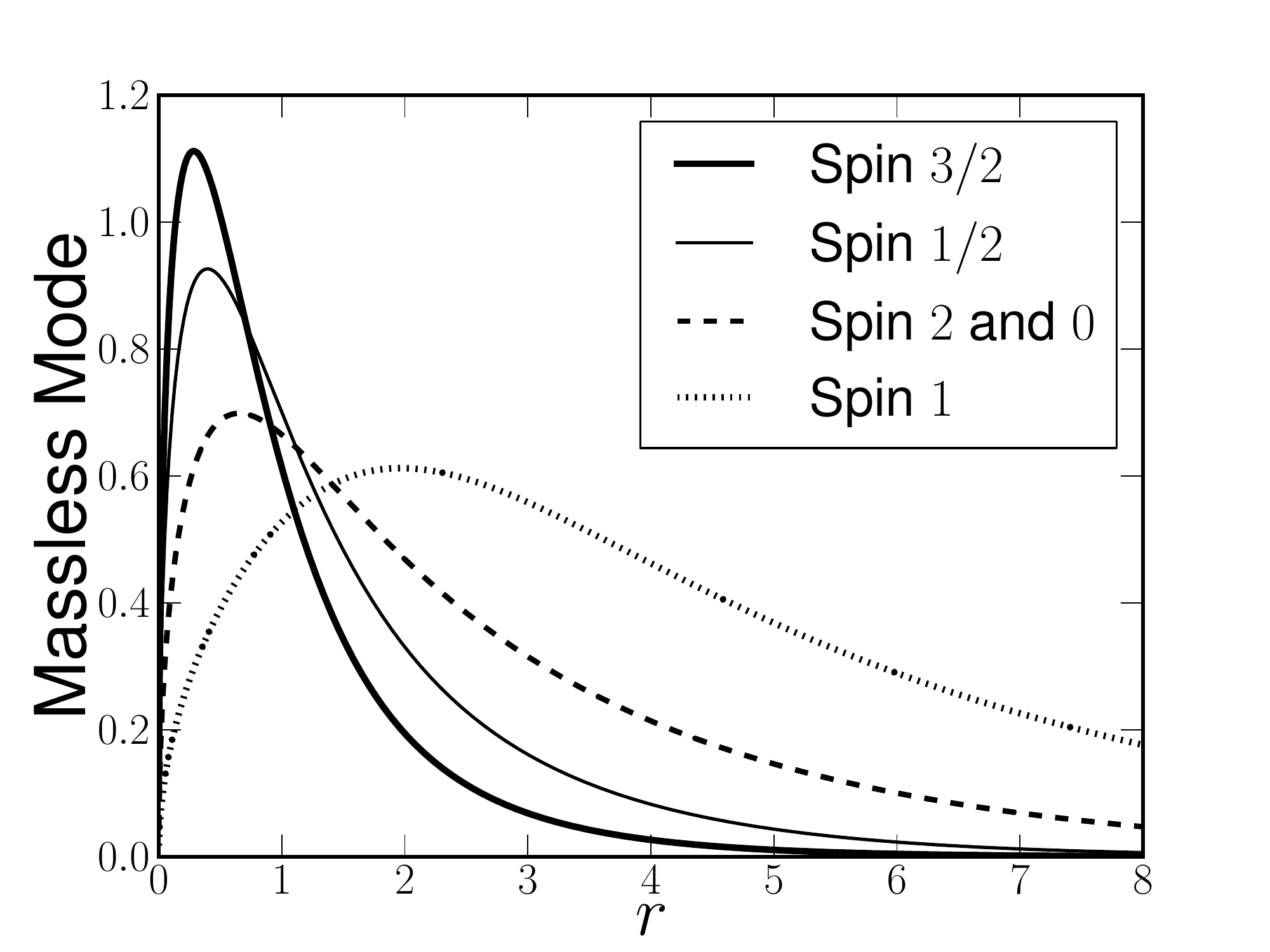}
                \caption{Comparative plot of the massless modes of bosonic and fermionic fields in the String-Cigar model for $c = 0.50$ and $\lambda=2.0$ (in the fermionic case). The derivative of the spinorial fields are indefinite at origin promoting similarity with the bosonic fields.}
                \label{Fig_Zero_Modes}
        \end{minipage}
        ~ 

\end{figure}

Now, we use the same $A_{\theta}$ fixed in Eq. \eqref{athetar}, which implies that the $\mathcal{W}(r)=-\lambda \mathcal{P}(r)$ of Eq. \eqref{qfix} holds. Then, in order to obtain a normalized squared modulus solution, the massless-mode of  \textit{rigth-handed} spin $3/2$ takes the form
\begin{eqnarray}
\label{uzero3/2}
u^0_{R}(r)=\mathcal{C}_0\exp\left[ \int_{r^{\prime}}{dr^{\prime}}\left(\left(\lambda-1\right)\mathcal{P}-\frac{F^{\prime}}{2F} \right)\right]=\mathcal{C}_0  F^{(\lambda-\frac{1}{2})}(r)H^{\frac{1}{4}(\lambda-1)}(r).
\end{eqnarray}
where $\mathcal{C}_0$ is a normalization constant. Afresh, for $\lambda=0$, the expression \eqref{uzero3/2} is the same obtained in Ref. \cite{Oda:2000zc} which is non-normalizable. Moreover, the radial effective action for spin $3/2$ is the same of the spin $1/2$ case presented in Eq. \eqref{anorm2}. Note that there is a correlation between the massless modes  in  Equations \eqref{uzero3/2} and \eqref{alphar0R} of the form
 \begin{eqnarray}
 u_{R,L}^{0}(r)=F^{-\frac{1}{2}}\alpha_{R,L}^{0}(r).
 \label{ua}
  \end{eqnarray} 
This change promotes a small increase of the amplitude of the zero mode for the spin $3/2$. We plot in the Figure \ref{Fig_MasslessMode-Rarita} the massless mode of the Rarita-Schwinger field \eqref{uzero3/2} and compare with Eq. \eqref{alphar0R}. 

At this point, we have the expressions of the massless modes for the gravitational and scalar fields in Eq. \eqref{Massless_Grav-Charuto}, $U(1)$ gauge field in \eqref{Massless_Gauge-Charuto}, spin $1/2$ fields in \eqref{alphar0R} and spin $3/2$ in \eqref{uzero3/2}. We verified that for $\lambda=2$ the zero mode of fermionic fields shares similar profile to the bosonic fields. A comparative plot is made in Figure \ref{Fig_Zero_Modes}.



Using the explicit expressions of the warp factors in Eq. \eqref{Eq_WarpFactors} for string-cigar, the zero mode  of spin $3/2$ field can be written as
\begin{equation}\label{u0R}
u^0_{R}(r) =\mathcal{C}_0\left(\frac{\tanh{(cr)}}{c}\right)^{\frac{(\lambda-1)}{2}}\exp{\left(\frac{1}{4}(3-5\lambda)[cr + \tanh{(cr)}]\right)}.
\end{equation}
Likewise the spin $1/2$ case, for the existing of a zero mode (which vanishes at the origin) is required that $\lambda>1$, but only for $\lambda>3$ there is a normalizable mode with its derivatives null at the origin. Moreover, for the \textit{left-handed} massless modes is required that $\lambda \to-\lambda$ and only one chiral mode is allowable to exist in the brane as well.

\subsection{Spin $3/2$ Massive Modes}\label{Sec_MassiveModes-3/2}
%
Using the conformal radial coordinate $z$ \eqref{Eq_z(r)}, we decouple the equation \eqref{zero3/2} in following two second order differential equations:
\begin{eqnarray}\label{stlz-3/2}
\left[\partial_z^2+2\left(\tilde{\mathcal{P}}(z)-\frac{\dot{F}}{2F}\right)\partial_z+\left\{(1\mp\lambda)\dot{\tilde{\mathcal{P}}}(z)-\partial_z\left(\frac{\dot{F}}{2F}\right)+\right.\right.\hspace{2cm}\nonumber\\
\left.\left.+(1-\lambda^2)\tilde{\mathcal{P}}^2(z)-\frac{\tilde{\mathcal{P}}(z)\dot{F}}{F}+\left(\frac{\dot{F}}{2F}\right)^2\right\}\right]u_{R,L}=- m^2 u_{R,L}(z),
\end{eqnarray}
where $\tilde{\mathcal{P}}(z)=\sqrt{F}P(z)$.

Returning to original variable $r$, the equation \eqref{stlz-3/2} turns to
\begin{eqnarray}\label{stlr-3/2}
\left[\partial_r^2+\left(2f+\frac{g}{2}\right)\partial_r+\left\{\frac{(1\mp\lambda)}{8}\Big[5f^2+fg +10f^{\prime} + 2g^{\prime}\Big]+(1-\lambda^2)\left[\frac{5}{4}f+\frac{g}{4}\right]^2+\right.\right.\hspace{2cm}\nonumber\\
\left.\left.-\left[\frac{5}{4}f^2+\frac{fg}{4}+\frac{f^{\prime}}{2}\right]\right\}\right]u_{R,L}(r)=- \frac{m^2}{F} u_{R,L}(r). \hspace{2cm}
\end{eqnarray}

\begin{figure}[htb]
 \centering
    \includegraphics[width=0.50\textwidth]{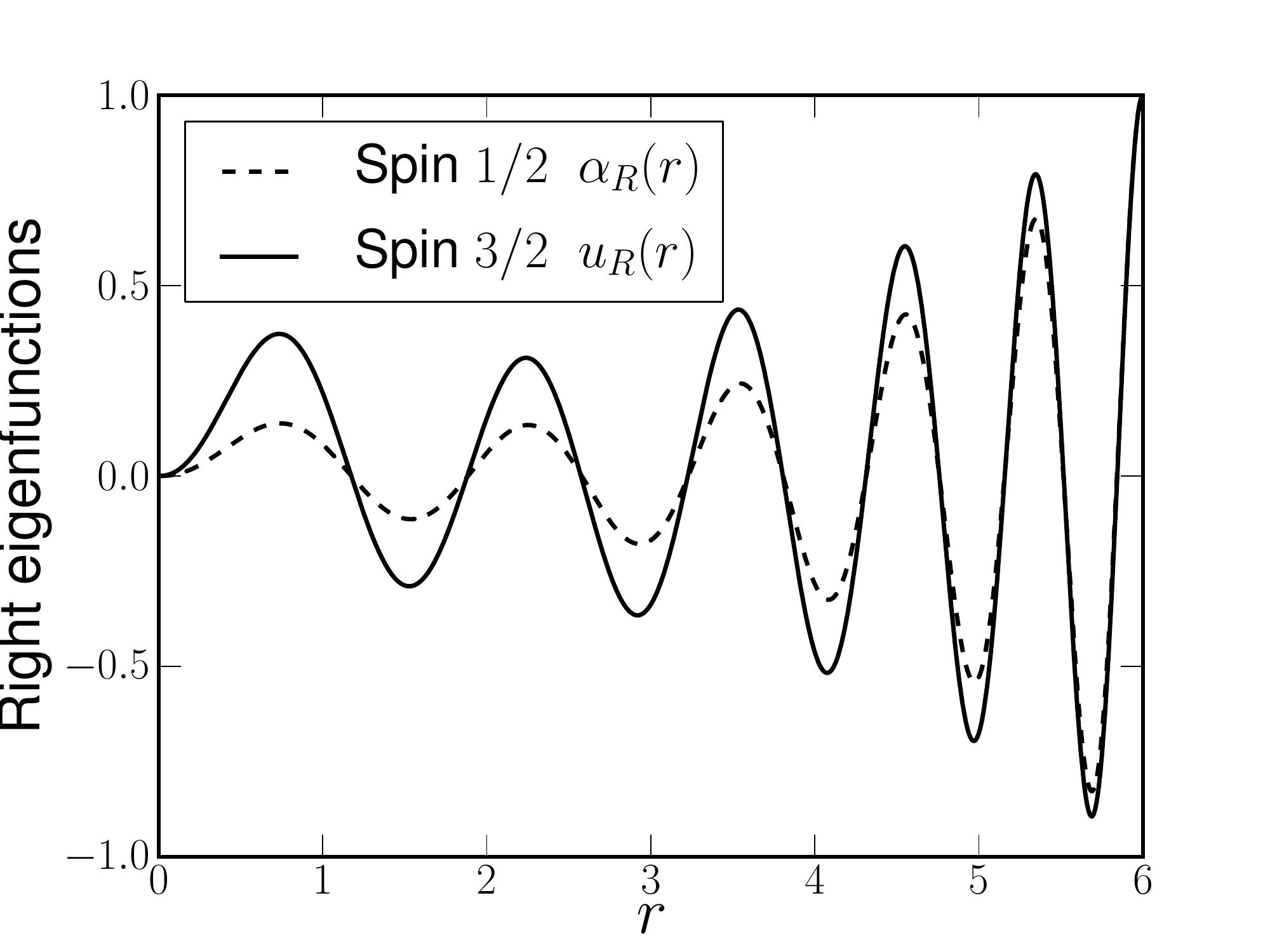}
 \caption{Comparative plot of  normalized right-handed eigenfunctions for spinorial fields for $\lambda = 5.0$ and $c=0.5$ in the String-Cigar model. The most closest mass eigenvalues values was found as $m_{1/2} = 0.4024$ and $m_{1/2} = 0.4016$.}
 \label{Fig_Autovetor-Rarita}
\end{figure}

Minor changes are noted comparing the expressions \eqref{stlr-3/2} and \eqref{Eq_SturmLiouville}. In the numerical solution of the Sturm-Liouville problem for the Rarita-Schwinger field  we verified that the mass spectrum is indistinguishable of that presented in Fig. \ref{Fig_Spectra}. Furthermore, the only difference arises in the eingefunctions. The amplitudes for the spin $3/2$ are higher than the spin $1/2$ case. This can be seen in the figure \ref{Fig_Autovetor-Rarita}. We intend to perform an analytical study of the spin $3/2$ massive modes in GS model in a future work.


\subsection{Schr\"{o}dinger approach for spin $3/2$}
Performing a change of the depend variable of  Eq. \eqref{stlz-3/2} in the form
\begin{equation}
u_{R,L}(z)=\exp \left[-\int_{z^{\prime}}{\left(\tilde{\mathcal{P}}(z)-\frac{\dot{F}}{2F}\right)}dz^{\prime}\right]\tilde{u}_{R,L}(z),
\label{Eq_SecondTransformation-3/2}
\end{equation}
in the Eq. \eqref{stlz-3/2}, we obtain a Schr\"odinger-like equation as
\begin{eqnarray}
\label{Eq_Schroedinger-3/2}
\left[-\partial_z^2 + V^{3/2}_{R,L}(z)\right]\tilde{u}_{R,L}(z)=m^2\tilde{u}_{R,L}(z),
\end{eqnarray}
where
\begin{equation}
\label{Eq_Potential-l3/2}
V^{3/2}_{R,L}(z)=\lambda^2\tilde{\mathcal{P}}^2(z)\mp\lambda\partial_z\tilde{\mathcal{P}}=
\tilde{\mathcal{W}}^2(z)\pm \partial_z\tilde{\mathcal{W}}.
\end{equation}

The spin $3/2$ Schr\"{o}dinger-like equation \eqref{Eq_Potential-l3/2} has the same expression of the spin $1/2$ field \eqref{Eq_Potential}, since the change of variables eliminates the multiplicative factor in the first order derivative term of $u_{R,L}$ in Eq.  \eqref{stlz-3/2}. As shown in section \ref{spin1/2schrodinger}, the form of the potential \eqref{Eq_Potential-l3/2} bears a SUSY-like symmetry that guarantees the equality of the KK massive spectra for both chiralities and 
the absence of tachyonic KK states.




\section{Conclusions and perspectives}
\label{conclusions}


In this work, we have studied the Dirac field in the 6D thick string-cigar braneworld. The string-cigar model is a thick string-like scenario which satisfies all the regularity and energy conditions. The thin string limit was also analyzed. We considered spin $1/2$ and $3/2$ (Rarita-Schwinger) bulk fields. We made a brief review of the localization of the gravitational and gauge bulk fields. We also studied the scalar field and we have shown that it has identical behaviour to that of the gravitational field \cite{Graviton_Charuto}.

A well-known feature of fermions in a string-like brane is that, unlike the gravitational, scalar and gauge fields, it is not possible to trap free fermions with a decreasing warp factor. Due to this fact, we proposed a suitable gauge coupling for the fermions with a background gauge field to confine both spin $1/2$ and spin $3/2$ fermions in the brane with positive tension.

Imposing suitable boundary conditions to guarantee the self-adjointness of the spinor operators, a normalized and everywhere well-defined massless mode is obtained for both the thin string and string-cigar models depending on the strength of the coupling constant $\lambda$. This is valid both for spin $1/2$ as for spin $3/2$ case. The massless modes have a shape similar to the energy-momentum components of the source of the string-cigar model, whose core is displaced from the origin. This shift of the core is a characteristic of string-like branes which source is a vortex for higher winding numbers \cite{Giovannini:2001hh}. Moreover, there is a tiny increase in the amplitude of the Rarita-Schwinger zero mode in comparison with spin $1/2$ field.

By means of numerical methods, we have obtained the spectrum and KK eigenfunctions. The absence of tachyonic states and the equality of the massive spectrum for the right and the left chiralities is ensured by the supersymmetric Quantum Mechanics structure of the analogue potential. For $m\ll c$, the spectrum exhibits the usual behaviour of the Kaluza-Klein theories. Further, we verified that the mass spectrum of the spin $1/2$ and $3/2$ fields are indistinguishable. For the massive modes, the gauge coupling constant plays a similar role with the bulk mass for 5D fermions \cite{gp}. The massive eigenfunctions behave like those of the thin string-like asymptotically. Near the core of the brane, the massive modes are enhanced. This behaviour is also present in the gravitation \cite{Graviton_Charuto} (and, consequently scalar) and vectorial \cite{Costa:2015dva} cases. Further, the amplitudes for the spin $3/2$ are higher than the spin $1/2$ case in whole domain.

Although the KK modes are not localized on the brane, some massive states may exhibit a resonant profile. The search of these states was performed by means of the resonance method. We found peaks in the probability distribution relating to states which wavefunctions have very high amplitudes near the brane. Apart this feature, the oscillation of the wavefunction must be as small as possible to characterize a resonant state. The numerical solutions of the Schr\"odinger-like equation showed that this occurs for a specific left-handed mode. Thus, only fermions with \textit{one chirality} is allowed to interact with the brane as a resonant state. An important result is that the coupling parameter allows the existence of a resonance peak. The geometric parameter $c$ determines the magnitude of the resonant mass which is consistent with the fact that it is related to the Planck scale cut-off. Besides, in the Schr\"{o}dinger approach, the spin $1/2$ and $3/2$ fields share an identical structure, which prevents tachyonic KK modes for both fields.

Finally, we made a comparative of the normalized massless modes for the bosonic and fermionic fields. For a specific value of the gauge coupling, all the field have similar shapes. 

For future works, we intend to study numerically the KK modes, as well as the resonant states, of the spin $1/2$  and $3/2$ fermions in a warped resolved conifold \cite{Dantas:2013iha}. This scenario provides a geometric flow that controls the singularity at the origin in the quantum analogue potential. Moreover, we intend to study the properties of gauge coupling term for $l \neq 0$ waves and its influence over the masseless and the KK modes.

\section{Acknowledgments}

The authors thank the Coordena\c{c}\~{a}o de Aperfei\c{c}oamento de Pessoal de N\'{i}vel Superior (CAPES), the Conselho Nacional de Desenvolvimento Cient\'{i}fico e Tecnol\'{o}gico (CNPQ), and Funda\c{c}\~{a}o Cearense de apoio ao Desenvolvimento Cient\'{i}fico e Tecnol\'{o}gico (FUNCAP) for financial support. D. M. Dantas thanks to Rold\~{a}o da Rocha and  Marcelo B. Hott for useful discussions. J. E. G. Silva acknowledges the Indiana University Center for Spacetime Symmetries for the kind hospitality.



\begin{thebibliography}{99}

\bibitem{ADD}   
	N. Arkani-Hamed, S. Dimopoulos and G. R. Dvali, Phys.\ Lett.\ B {\bf 429}, 263 (1998).

\bibitem{rs}
    L.~Randall and R.~Sundrum, Phys.\ Rev.\ Lett.\  {\bf 83}, 3370 (1999); 
    L.~Randall and R.~Sundrum, Phys.\ Rev.\ Lett.\  {\bf 83}, 4690 (1999). 

\bibitem{merab}
M.~Gogberashvili,
  Mod.\ Phys.\ Lett.\ A {\bf 14}, 2025 (1999).

\bibitem{warpedmetrics} 
	V. A. Rubakov and M. E. Shaposhnikov, Phys.\ Lett.\ B {\bf 125}, 136 (1983).
	V.~A.~Rubakov and M.~E.~Shaposhnikov, Phys.\ Lett.\ B {\bf 125}, 139 (1983).
  C.~Wetterich,
  Nucl.\ Phys.\ B {\bf 242}, 473 (1984).
%
	M. Visser, Phys. Lett. B \textbf{159}, 22 (1985).
  M.~Gell-Mann and B.~Zwiebach,
  Nucl.\ Phys.\ B {\bf 260}, 569 (1985).


\bibitem{cosmologicalconstant}
N.~Arkani-Hamed, S.~Dimopoulos, N.~Kaloper and R.~Sundrum,
  Phys.\ Lett.\ B {\bf 480}, 193 (2000). J.~-W.~Chen, M.~A.~Luty and E.~Ponton,
  JHEP {\bf 0009}, 012 (2000). T.~Koivisto, D.~Wills and I.~Zavala,
  JCAP {\bf 1406}, 036 (2014).
  H.~M.~Lee and A.~Papazoglou,
  Phys.\ Rev.\ D {\bf 80}, 043506 (2009). J.~M.~Schwindt and C.~Wetterich,
  Nucl.\ Phys.\ B {\bf 726}, 75 (2005).

  \bibitem{darkmatter}
  G.~Panico, E.~Ponton, J.~Santiago and M.~Serone,
  Phys.\ Rev.\ D {\bf 77}, 115012 (2008)
  T.~Gherghetta and B.~von Harling,
  JHEP {\bf 1004}, 039 (2010)
  A.~D.~Medina and E.~Ponton,
  JHEP {\bf 1109}, 016 (2011).



\bibitem{rsfields}
  W.~D.~Goldberger and M.~B.~Wise,
  Phys.\ Rev.\ Lett.\  {\bf 83}, 4922 (1999)
  O.~DeWolfe, D.~Z.~Freedman, S.~S.~Gubser and A.~Karch,
  Phys.\ Rev.\ D {\bf 62}, 046008 (2000).
	Martin Gremm, Phys. Lett. B \textbf{478}, 434-438 (2000).
  D.~Bazeia and A.~R.~Gomes,
  JHEP {\bf 0405}, 012 (2004).


\bibitem{Gauge-5D}
	B.~Bajc and Gabadadze, Phy. Lett. B {\bf 474}, 282-291 (2000). 


\bibitem{Kehagias:2000au}
  A.~Kehagias and K.~Tamvakis,
Phys.\ Lett.\ B {\bf 504}, 38 (2001).

\bibitem{Csaki1}   
    C.~Csaki, J.~Erlich, T.~J.~Hollowood and Y.~Shirman, Nucl.\ Phys.\ B {\bf 581}, 309 (2000).

\bibitem{Csaki2} 
    C. Csaki, J. Erlich, and T. J. Hollowood, Phys. Rev. Lett. \textbf{84}, 5932 (2000).

\bibitem{CASA-Fermion-TwoField-ThickBrane} 
    C.~A.~S.~Almeida, M.~M.~Ferreira, Jr., A.~R.~Gomes and R.~Casana, Phys.\ Rev.\ D {\bf 79}, 125022 (2009).

\bibitem{Chineses-Ressonance} 
  	Yu-Xiao Liu, Jie Yang, Zhen-Hua Zhao, Chun-E Fu, and Yi-Shi Duan, Phys. Rev. D \textbf{80}, 065019 (2009).


\bibitem{Cruz:2011kj}
  W.~T.~Cruz, A.~R.~Gomes and C.~A.~S.~Almeida,
  Europhys.\ Lett.\  {\bf 96}, 31001 (2011)


\bibitem{German:2012rv}
  G.~German, A.~Herrera-Aguilar, D.~Malagon-Morejon, R.~R.~Mora-Luna and R.~da Rocha,
  JCAP {\bf 1302}, 035 (2013)

	\bibitem{Gauge-5D-Dilaton}
	W.~T.~Cruz, M.~O.~Tahim and C.~A.~S.~Almeida, Phys.\ Lett.\ B {\bf 686}, 259-263 (2010); 
	W.~T.~Cruz, A.~R.~P.~Lima and C.~A.~S.~Almeida, Phy. Rev. D {\bf 87}, 045018 (2013). 

\bibitem{Tofighi:2014vaa}
  A.~Tofighi and M.~Moazzen,
  Int.\ J.\ Mod.\ Phys.\ A {\bf 29}, no. 24, 1450126 (2014).

\bibitem{Chumbes:2011zt}
  A.~E.~R.~Chumbes, J.~M.~Hoff da Silva and M.~B.~Hott,
  Phys.\ Rev.\ D {\bf 85}, 085003 (2012)

  	
\bibitem{Wilami1}  
W.~T.~Cruz, A.~R.~Gomes and C.~A.~S.~Almeida, Eur. Phy. J. C \textbf{71}, 1790 (2011)

\bibitem{Correa:2010zg} 
    R.~A.~C.~Correa, A.~de Souza Dutra and M.~B.~Hott, Class.\ Quant.\ Grav.\  {\bf 28}, 155012 (2011).




\bibitem{Castro:2010uj} 
    L.~B.~Castro and L.~A.~Meza, Europhys.\ Lett.\  {\bf 102}, 21001 (2013);   L.~B.~Castro,
  Phys.\ Rev.\ D {\bf 83}, 045002 (2011)
  H.~Guo, Q.~Y.~Xie and C.~E.~Fu, arXiv:1408.6155 [hep-th].


\bibitem{Rahman:2011ik}
  R.~Rahman,
  Phys.\ Rev.\ D {\bf 87}, no. 6, 065030 (2013)

\bibitem{torsion}
  B.~Mukhopadhyaya, S.~Sen and S.~SenGupta,
  Phys.\ Rev.\ Lett.\  {\bf 89}, 121101 (2002).  N.~R.~F.~Braga and C.~N.~Ferreira,
  JHEP {\bf 0503}, 039 (2005).
  J.~M.~Hoff da Silva and R.~da Rocha,
  Phys.\ Rev.\ D {\bf 81}, 024021 (2010)
J.~Yang, Y.~-L.~Li, Y.~Zhong and Y.~Li,
  Phys.\ Rev.\ D {\bf 85}, 084033 (2012).


\bibitem{weylgeometry}
  N.~Barbosa-Cendejas and A.~Herrera-Aguilar,
  Phys.\ Rev.\ D {\bf 73}, 084022 (2006).
  Y.~-X.~Liu, X.~-H.~Zhang, L.~-D.~Zhang and Y.~-S.~Duan,
  JHEP {\bf 0802}, 067 (2008).  H.~Guo, A.~Herrera-Aguilar, Y.~X.~Liu, D.~Malagon-Morejon and R.~R.~Mora-Luna,
  Phys.\ Rev.\ D {\bf 87} (2013) 9,  095011.

\bibitem{cosmicstring}
W. Israel, Phys. Rev. D \textbf{15}, 935 (1977).
A. Vilenkin, Phys. Rev. Lett. \textbf{46}, 17 1169 (1981);
 R. Gregory, Phys. Rev. Lett. \textbf{59}, 740 (1987);
A. G. Cohen and D. B. Kaplan,  Phys.\ Lett.\ B {\bf 215}, 67 (1988);
V. P. Frolov, W. Israel and W. G. Unruh, Phys. Rev. D \textbf{39}, 1084 (1989);
M. Christensen, A. L. Larsen and Y. Verbin, Phys. Rev. D \textbf{60}, 125012 (1999).

\bibitem{stringlike}
    A.~G.~Cohen and D.~B.~Kaplan, Phys.\ Lett.\  B {\bf 470}, 52 (1999);  
    I.~Olasagasti and A.~Vilenkin, Phys.\ Rev.\ D {\bf 62}, 044014 (2000).   
 A.~Chodos and E.~Poppitz,
  Phys.\ Lett.\ B {\bf 471}, 119 (1999).

\bibitem{gregory}
R.~Gregory, Phys.\ Rev.\ Lett.\  {\bf 84}, 2564 (2000).

\bibitem{Gherghetta:2000qi}  
    T.~Gherghetta and M.~E.~Shaposhnikov, Phys.\ Rev.\ Lett.\  {\bf 85}, 240 (2000).

\bibitem{ponton}
  E. Ponton and E. Poppitz, JHEP {\bf 0102}, 042 (2001).




\bibitem{Giovannini:2001hh} 
    M.~Giovannini, H.~Meyer and M.~E.~Shaposhnikov,  Nucl.\ Phys.\ B {\bf 619}, 615 (2001).

\bibitem{Oda-PRD} 
I.\,Oda, Phys. Rev. D \textbf{62}, 126009 (2000).

    	
\bibitem{Oda:2000zc}   
    I.~Oda, Phys.\ Lett.\  B {\bf 496}, 113 (2000).

\bibitem{Oda:2003jc}
  I.~Oda,
  Phys.\ Lett.\ B {\bf 571}, 235 (2003).

\bibitem{Liu:2007gk}  
    Y.~-X.~Liu, L.~Zhao and Y.~-S.~Duan, JHEP {\bf 0704}, 097 (2007).

\bibitem{Liu:2007eb} 
  Y.~X.~Liu, L.~Zhao, X.~H.~Zhang and Y.~S.~Duan,
  Nucl.\ Phys.\ B {\bf 785}, 234 (2007)
    

\bibitem{Giovannini_Gauge-6D-PRD}
 M.~Giovannini,
  Phys.\ Rev.\ D {\bf 66}, 044016 (2002).


\bibitem{Giovannini_Gauge-6D-CQG}
M.~Giovannini, J.~V.~Le Be and S.~Riederer,
  Class.\ Quant.\ Grav.\  {\bf 19}, 3357 (2002).


\bibitem{Giovannini-5D}
M.~Giovannini,
  Phys.\ Rev.\ D {\bf 65}, 124019 (2002).


\bibitem{luis}
L.~J.~S.~Sousa, W.~T.~Cruz and C.~A.~S.~Almeida,
  Phys.\ Lett.\ B {\bf 711}, 97 (2012).

\bibitem{tinyakov}
  P.~Tinyakov and K.~Zuleta,
  Phys.\ Rev.\ D {\bf 64}, 025022 (2001).

\bibitem{codimension2gravity}
  I.~Navarro and J.~Santiago,
  JHEP {\bf 0502}, 007 (2005)
I.~Navarro,
  JCAP {\bf 0309}, 004 (2003)
  I.~Navarro and J.~Santiago,
  JHEP {\bf 0404}, 062 (2004)
  I.~Navarro and J.~Santiago,
  JCAP {\bf 0603}, 015 (2006)
 S.~Kanno and J.~Soda,
  JCAP {\bf 0407}, 002 (2004)
 G.~Kofinas,
  Phys.\ Lett.\ B {\bf 633}, 141 (2006)
  P.~Bostock, R.~Gregory, I.~Navarro and J.~Santiago,
  Phys.\ Rev.\ Lett.\  {\bf 92}, 221601 (2004).


\bibitem{codimension2cosmology}
  J.~M.~Cline, J.~Descheneau, M.~Giovannini and J.~Vinet,
  JHEP {\bf 0306}, 048 (2003). J.~Vinet and J.~M.~Cline,
  Phys.\ Rev.\ D {\bf 70}, 083514 (2004)
  JCAP {\bf 0507}, 004 (2005)
  E.~Papantonopoulos and A.~Papazoglou,
  JHEP {\bf 0509}, 012 (2005)
  E.~Papantonopoulos, A.~Papazoglou and M.~Tsoukalas,
  Phys.\ Rev.\ D {\bf 84}, 025016 (2011).
   C.~Charmousis and R.~Zegers,
  JHEP {\bf 0508}, 075 (2005).
  F.~Niedermann, R.~Schneider, S.~Hofmann and J.~Khoury,
  Phys.\ Rev.\ D {\bf 91}, no. 2, 024002 (2015).


\bibitem{cigaruniverse}
  B. de Carlos and J. M. Moreno, 
  JHEP {\bf 0311}, 040 (2003).

\bibitem{teardrop}
  A. Kehagias, Phys.\ Lett.\  B {\bf 600}, 133 (2004).

\bibitem{conifold}     J.~E.~G.~Silva and C.~A.~S.~Almeida, Phys.\ Rev.\ D {\bf 84}, 085027 (2011).   

\bibitem{Costa:2013eua}
  F.~W.~V.~Costa, J.~E.~G.~Silva and C.~A.~S.~Almeida,
  Phys.\ Rev.\ D {\bf 87}, 125010 (2013).


  \bibitem{einsteinmanifolds}
  S. Randjbar-Daemi and M. E. Shaposhnikov,
  Phys.\ Lett.\ B {\bf 491}, 329 (2000).
    C.~Wetterich,
  Phys.\ Rev.\ D {\bf 78}, 043503 (2008).


\bibitem{apple}
  M. Gogberashvili, P. Midodashvili and D. Singleton, JHEP {\bf 0708}, 033 (2007).

 \bibitem{football}
  J. Garriga and M. Porrati, JHEP {\bf 0408}, 028 (2004).
  C.~P.~Burgess, C.~de Rham, D.~Hoover, D.~Mason and A.~J.~Tolley,
  JCAP {\bf 0702}, 009 (2007).



\bibitem{smoothed}
  Y.~-S.~Duan, Y.~-X.~Liu and Y.~-Q.~Wang,
  Mod.\ Phys.\ Lett.\ A {\bf 21}, 2019 (2006).
J.~C.~B.~Ara\'{u}jo, J.~E.~G.~Silva, D.~F.~S.~Veras and C.~A.~S.~Almeida 
    Eur. Phys. J. C \textbf{75}, 127 (2015).
    
    
\bibitem{Silva:2012yj}  
    J.~E.~G.~Silva, V.~Santos and C.~A.~S.~Almeida, Class.\ Quant.\ Grav.\  {\bf 30}, 025005 (2013).

\bibitem{chow} B. Chow, P. Lu, L. Ni, Hamilton's Ricci flow, Science Press, 2006.

\bibitem{Friedan}
  D.~Friedan,
  Phys.\ Rev.\ Lett.\  {\bf 45}, 1057 (1980). A. A. Tseytlin, 
	Phys. Rev. D \textbf{75}, 064024 (2007). T. Oliynyk, V. Suneeta and E. Woolgar, 
	Phys. Rev. D {\bf 76}, 045001 (2007).

\bibitem{Lashkari:2010iy}
N. Lashkari and A. Maloney, 
	Class. Quant. Grav.  {\bf 28}, 105007 (2011).

\bibitem{Orth:2012ri}
P. P. Orth, P. Chandra, P. Coleman and J. Schmalian,
	Phys.\ Rev.\ Lett.\  {\bf 109}, 237205 (2012).


\bibitem{Graviton_Charuto} 
    D.\,F.\,S. Veras, J.\,E.\,G. Silva, W.\,T. Cruz and C.\,A.\,S. Almeida, Phys. Rev. D \textbf{91}, 065031 (2015).

\bibitem{Costa:2015dva} 
	F.~W.~V.~Costa, J.~E.~G.~Silva, D.~F.~S.~Veras and C.~A.~S.~Almeida, Phys.\ Lett.\ B {\bf 747}, 517 (2015)

\bibitem{Panotopoulos:2007fg} 
  G.~Panotopoulos,
  JCAP {\bf 0705}, 016 (2007)


\bibitem{Parameswaran:2006db} 
    S.~L.~Parameswaran, S.~Randjbar-Daemi and A.~Salvio, Nucl.\ Phys.\ B {\bf 767}, 54 (2007).

\bibitem{Dantas:2013iha} 
    D.~M.~Dantas, J.~E.~G.~Silva and C.~A.~S.~Almeida, Phys.\ Lett.\  B {\bf 725}, 425 (2013);

\bibitem{Sousa:2014dpa} 
   L.~J.~S.~Sousa, C.~A.~S.~Silva, D.~M.~Dantas and C.~A.~S.~Almeida,  Phys.\ Lett.\ B {\bf 731}, 64 (2014).


\bibitem{Budinich:2001nh}  
    P.~Budinich, Found.\ Phys.\  {\bf 32}, 1347 (2002).







\bibitem{MatrixMethod} 
	Pierluigi Amodio and Giuseppina Settanni, J. Numer. Anal. Indust. Appl. Math vol.\textbf{6} no. 1-2, 2011, pp.1-13.

\bibitem{gp}
    T.~Gherghetta and A.~Pomarol, Nucl.\ Phys.\ B {\bf 586}, 141 (2000); 
    Nucl.\ Phys.\ B {\bf 602}, 3 (2001); 
    S.~J.~Huber and Q.~Shafi, Phys.\ Lett.\ B {\bf 498}, 256 (2001); 
    Y.~Grossman and M.~Neubert, Phys.\ Lett.\ B {\bf 474}, 361 (2000).   

\bibitem{susy}
    F.~Cooper and B.~Freedman, Annals Phys.\  {\bf 146}, 262 (1983);  
    C.~V.~Sukumar, J.\ Phys.\ A {\bf 18}, 2917 (1985); 
    F.~Cooper, A.~Khare and U.~Sukhatme, Physics Reports \textbf{251}, (1995) 267-385.  

\bibitem{Numerov}
	B.V. Numerov, Monthly Notices of the Royal Astronomical Society, Vol. \textbf{84}, p.592-592 (1924); 
	B.V. Numerov, Astronomische Nachrichten, Vol. \textbf{230}, p.359 (1927). 





\bibitem{watson}
    G. Watson, A treatise on the theory of the Bessel functions, Cambridge University Press, 1996.






\end{thebibliography}
\end{document}